\documentclass[twocolumn,preprintnumbers,amsmath,amssymb,10pt,a4paper,showpacs,superscriptaddress,nofootinbib]{revtex4}
\usepackage{geometry}
\usepackage{graphicx}
\usepackage{dcolumn}
\usepackage{bm}

\newcommand{\plus}{\!+\!}
\newcommand{\minus}{\!-\!}
\newcommand{\equals}{\!=\!}
\newcommand{\timess}{\!\times\!}
\newcommand{\les}{\!\le\!}
\newcommand{\lt}{\!<\!}

\newcommand{\gt}{\!>\!}

\renewcommand{\vec}[1]{\mbox{\boldmath$#1$}}
\newcommand{\vecs}[1]{\mbox{\scriptsize\boldmath$#1$}}

\newcommand{\avg}[1]{\left\langle{#1}\right\rangle}

\newcommand{\beq}{\begin{equation}}
\newcommand{\eeq}{\end{equation}}
\newcommand{\bal}{\begin{aligned}}
\newcommand{\eal}{\end{aligned}}

\newcommand{\be}{\begin{equation}}
\newcommand{\ee}{\end{equation}}
\newcommand{\bd}{\begin{displaymath}}
\newcommand{\ed}{\end{displaymath}}
\newcommand{\BE}{\begin{eqnarray}}
\newcommand{\EE}{\end{eqnarray}}

\begin{document} 

\title{Pattern formation in individual-based systems with time-varying parameters}

\author{Peter Ashcroft}
\email{peter.ashcroft@postgrad.manchester.ac.uk}
\affiliation{Theoretical Physics, School of Physics and Astronomy, The University of Manchester, Manchester M13 9PL, United Kingdom}

\author{Tobias Galla}
\email{tobias.galla@manchester.ac.uk}
\affiliation{Theoretical Physics, School of Physics and Astronomy, The University of Manchester, Manchester M13 9PL, United Kingdom}

\begin{abstract} 
We study the patterns generated in finite-time sweeps across symmetry-breaking bifurcations in individual-based models. Similar to the well-known Kibble-Zurek scenario of defect formation, large-scale patterns are generated when model parameters are varied slowly, whereas fast sweeps produce a large number of small domains. The symmetry breaking is triggered by intrinsic noise, originating from the discrete dynamics at the micro-level. Based on a linear-noise approximation, we calculate the characteristic length scale of these patterns. We demonstrate the applicability of this approach in a simple model of opinion dynamics, a model in evolutionary game theory with a time-dependent fitness structure, and a model of cell differentiation. Our theoretical estimates are confirmed in simulations. In further numerical work, we observe a similar phenomenon when the symmetry-breaking bifurcation is triggered by population growth. 
\end{abstract}

\maketitle
\section{Introduction}
The systematic analytical description of pattern-forming systems started with the seminal work by Alan Turing in the 1950's \cite{AT52}, and it has found numerous applications in chemistry, fluid dynamics, biology, and other disciplines \cite{CG09}. Turing's work initiated a large systematic effort to characterize spatial dynamical systems, their attractors, and instabilities. A systematic classification of spatial instabilities is now available in the literature of non-linear dynamics \cite{CG09}, and the resulting patterns are well understood. The precise mechanisms and reactants at work on the microscopic level are still being unearthed in many biological systems, and while it has become clear that Turing's ideas are not applicable to all systems for which they were initially developed, his picture of pattern-forming systems is still one of the cornerstones of modern dynamical systems theory. 

While Turing's theory describes partial differential equations at constant parameters, a related, but separate picture of defect formation was developed first in the theory of cosmological systems \cite{Kibble76}, and later in condensed matter physics \cite{Zurek85}. It is now known as the Kibble-Zurek (KZ) theory of defect formation, and it describes situations in which a system is swept through an instability slowly in finite time. This out-of-equilibrium process sets the density of defects separating domains of constant order parameter \cite{LZ97}. The typical length scale separating the defects, and hence the scale of the resulting pattern, is determined by the quench rate; slower sweeps result in patterns with large length scales, fast quenches lead to a larger number of domain walls or other topological defects, see e.g. \cite{lythe,LZ97,YZ98,ML99,GM03} for theoretical approaches, or \cite{LCD,LCD2,Neutron,Neutron2,DR99,CL12,hendry,ulm,casado} for experimental results.   

The starting point for the theoretical analysis of both the formation of Turing patterns and for the KZ picture of defect creations is models defined by partial differential equations. These models are formulated in terms of continuous order-parameter fields, describing, for example, the concentrations of chemicals or biological agents, or objects such as gauge fields or particle densities in the context of cosmology \cite{Kibble76,Zurek85,AT52,CG09,LZ97,YZ98,lythe,ML99,GM03}. It is only more recently that pattern-forming processes have been considered in individual-based systems. Here, the dynamics are described in terms of reactions of discrete particles. In chemistry for example, a molecule of one type may react with a molecule of a different type, in ecology a predator and a prey individual interact, and in cell biology a progenitor cell may `decide' to develop into one of several cell types. These systems all evolve through a series of transitions between discrete states, occurring as stochastic processes with specified transition rates. A description in terms of deterministic differential equations is here only adequate if the number of particles in the system is large (formally infinite) and when stochastic effects can be neglected. In finite systems the stochasticity of the underlying discrete-particle dynamics can give rise to relevant effects not captured by the deterministic limiting description. For example, this so-called demographic stochasticity has been seen to induce persistent cycles in a variety of models \cite{MN05,RMF06,KGG07,PBD07,AMP07,STN08}. Spatial systems subject to demographic noise can develop patterns and traveling waves in parameter regimes in which a purely deterministic description would predict a stable spatially uniform fixed point \cite{BG11,BF10}.  These phenomena are known as quasi-cycles, quasi-Turing patterns and traveling quasi-waves, and they can be described and predicted analytically in the so-called linear-noise approximation \cite{LM08,SPT11,BGM11}.

These pattern- and wave-forming phenomena, and their analytical description, constitute an extension of Turing's theory to individual-based systems. While Turing's approach relies on methods from non-linear dynamics, the analysis of noise-driven quasi-patterns combines these techniques with tools from non-equilibrium statistical physics. Both approaches are usually applied to systems with constant parameters, this is where Turing's theory applies and where quasi-Turing patterns are observed as well.

The purpose of the present work is to develop a similar picture for spatial individual-based systems which are swept across a Turing instability or symmetry-breaking bifurcation in finite time. In such systems one or more control parameters are time-dependent. For example, populations in fluctuating time-dependent environments are studied in \cite{leibler1, leibler2}, cellular decision making with time-varying external signals is investigated in \cite{alexei1,alexei2}, and models of population genetics with time-dependent selection pressure are considered in \cite{gerland}. Waddington's picture of a marble rolling down an epigenetic landscape with bifurcating valleys \cite{waddington} is another example of a system which is swept across a symmetry-breaking transition.

In the first part of this paper we consider stochastic interacting particle systems in a spatial setting in which one control parameter is increased linearly in time, moving the systems from a spatially homogeneous state into a symmetry-broken regime. The symmetry breaking is triggered by intrinsic fluctuations and leads to the formation of spatial domains of a characteristic length scale which is determined by the quench rate. Analytical predictions are possible within the linear-noise approximation, and we test these predictions against numerical simulations. Specifically, we investigate a simple model of opinion dynamics, chosen because it constitutes a microscopic realization of the Ginzburg-Landau equation, one of the most basic models with a symmetry-breaking transition. As a second example, we study a model of evolutionary dynamics in which the underlying fitness landscape changes with time. This could happen, for example, as a consequence of varying external factors. The third example finally is a model of decision making in cells. 

In the last part of the paper we consider a separate noise-driven mechanism of pattern formation. Growing populations subject to bistable dynamics are exposed to large noise levels at the beginning of the dynamics, when particle numbers are low. As the growth continues, noise levels are reduced and spatial structures form. As we show, the typical spatial extension of these domains scales with the growth rate.

The remainder of the paper is organized as follows: In Sec. \ref{sec:mech} we describe the basic mechanism underlying the pattern-forming process in systems with time-varying parameters. Sec. \ref{sec:opinion} contains an application to a model of opinion dynamics, recently studied by Russell and Blythe for fixed model parameters \cite{RB11}. In Sec. \ref{sec:evol} we then apply these ideas to a replicator-mutator system in evolutionary dynamics, before we discuss a model of decision making in cells in Sec. \ref{sec:cells}. Growing populations are considered in Sec. \ref{sec:growing}, and we summarise our findings and draw conclusions in Sec. \ref{sec:concl}.

\section{The basic mechanism: pattern formation in slow quenches}\label{sec:mech}
In this section we will briefly summarize the phenomenological picture underlying the KZ theory \cite{Kibble76,Zurek85}. For recent and in-depth investigations into the KZ theory see e.g. \cite{antunes06,biroli10,chandran12}. The theory describes systems with a spatially varying order parameter, say $\phi(x,t)$, and which can experience either an ordered or a disordered phase at equilibrium. Which phase the system is in is determined by a control parameter, $g$, which can for example represent a (reduced) temperature. For a general case we say that the two regimes are separated by a symmetry-breaking phase transition at ${g \equals g_b}$, where the subscript `$b$' indicates a bifurcation point of the dynamics. In our convention the disordered phase is the one in which ${g \lt g_b}$, and the ordered phase is the one in which ${g \gt g_b}$. In the vicinity of the transition, the dynamical relaxation time, $\tau$, and the correlation length, $\xi$, of the order parameter simultaneously diverge,
\begin{equation}
 \tau \sim |g-g_b|^{-\kappa}, \quad \xi \sim |g-g_b|^{-\nu}, \label{eq:mech_diverge}
\end{equation}
with exponents $\kappa$ and $\nu$ specific to the model at hand.

If the control parameter is swept through the phase transition as a linear function of time, ${g(t) \equals \mu t}$ (${\mu \gt 0}$), starting from the disordered phase ${g \lt g_b}$, the dynamic relaxation time will exceed the time scale on which the control parameter varies near ${g \equals g_b}$ and the system will cease to evolve adiabatically. Zurek \cite{Zurek85} estimated that this will happen at the time, $t_c$, when ${\dot{g}/(g \minus g_b) \!\approx\! \tau^{-1}}$, which directly leads to ${\hat{t} \!\approx\! \tau(t_c)}$, where ${\hat{t} \equals t_c \minus t_b}$ and where $t_b$ is the time at which ${g(t) \equals g_b}$. We denote quantities at the point where the system ceases to follow quasi-equilibrium by subscript $c$, and the distance between the bifurcation point and the critical point by hats (${\hat{t} \equals t_c\minus t_b}$, ${\hat{g} \equals g_c \minus g_b}$). One keeps in mind that the relaxation time and correlation length are functions of $g$, so that in our linear annealing protocol they are functions of $t$. In-line with the existing literature \cite{Zurek85} we will refer to the time $\hat{t}$ as the `freeze-out' time, this is the time that elapses between crossing the equilibrium bifurcation point, $g_b$, and reaching the point $g_c$ at which the system resumes its adiabatic motion. This reflects the observation that order-parameter fields of systems undergoing a quench frequently remain close to the equilibrium point in the disordered phase, even beyond the point where the sweep has progressed into the ordered regime, ${g \gt g_b}$. It is only after some delay that the system falls out of the unstable disordered equilibrium and that the order parameter assumes values typical for the ordered phase. Within the KZ picture, topological defects are created at this time, and then remain frozen into the dynamics, although some slow coarsening may follow. Details can be found in \cite{LZ97,YZ98,lythe,ML99,GM03}; we will occasionally refer to this phenomenon as a `delayed bifurcation' \cite{antunes06,biroli10,remark:defects}.

Using the scaling of Eq. (\ref{eq:mech_diverge}) one finds ${\hat{t} \!\sim\! \mu^{-\frac{\kappa}{1+\kappa}}}$, or equivalently, using ${\hat{g} \equals \mu\hat{t}}$,
\begin{equation}
\hat{g} \sim \mu^{\frac{1}{1+\kappa}} \label{eq:mech_ghat}.
\end{equation}
The length scale setting the density of defects is in turn given by 
\begin{equation}
 \hat{\xi} \sim \mu^{-\frac{\nu}{1+\kappa}} \label{eq:mech_zeros}.
\end{equation}
The scaling of the resulting length scale, $\hat{\xi}$, with the quench rate, $\mu$, depends on the equilibrium exponents $\kappa$ and $\nu$ of the system. For simple Ginzburg-Landau systems we have ${\nu \equals 1/2}$ and ${\kappa \equals 1}$ for example \cite{Zurek85}, so that the length scale follows ${\hat{\xi} \!\sim\! \mu^{-1/4}}$. In a one-dimensional system the number of defects created in a slow quench hence grows as $\mu^{1/4}$. In a two-dimensional system with point defects only the density of defects would scale as ${\hat{\xi}^{-2}}$, i.e. as the square root of the quench rate. These results have been verified in numerical simulations for a variety of different systems, and they are corroborated by analytical calculations based on linear approximations of the underlying stochastic partial differential equations \cite{LZ97,YZ98,ML99,GM03,lythe,DS01,D05}.

\section{Model of opinion dynamics}\label{sec:opinion}
As a first example of an individual-based system in which patterns are generated during a slow quench we consider a microscopic realization of the time-dependent Ginzburg-Landau equation. This equation is the archetypal example of a symmetry-breaking phase transition, and an individual-based realization has recently been proposed by Russell and Blythe \cite{RB11}. Specifically the authors consider a model of processes in linguistics; each individual can be in one of two discrete states, representing `two ways of saying the same thing' \cite{RB11}.  Similar models have been used in the context of opinion dynamics, where the two states represent two possible views any individual may have on a given subject (see Ref. \cite{castellano} for a review). The agents' choice of state is determined through interactions with other individuals in their immediate neighbourhood, potentially subject to a systematic bias towards one of the two states. We will detail this below. It should be kept in mind that this model is not specifically designed to model any real-world process, we here study it primarily as a microscopic realization of the Ginzburg-Landau dynamics. Any reference to this model as a model of opinion dynamics or of language dynamics is therefore metaphorical.

\subsection{Model definition}
The model is defined on a one-dimensional periodic lattice with $L$ sites, each hosting $N$ individuals (also referred to as spins). Sites will be labeled by ${\ell \equals 1,\dots,L}$, and each of the $NL$ spins can be in one of two states, up and down. We will write $n_\ell$ for the (time-dependent) number of up-spins in site $\ell$, and consequently the number of down-spins in site $\ell$ is ${N \minus n_\ell}$.  We will consider a continuous-time dynamics defined by the following rates:
\begin{eqnarray}
 T_{1,\ell}(n_\ell \minus 1|n_\ell) \!&=&\! (1\minus2D)\bigl[1\minus\Pi(n_\ell,t)\bigl]n_\ell, \nonumber\\
 T_{2,\ell}(n_\ell \plus 1|n_\ell)  \!&=&\! (1\minus2D)\Pi(n_\ell,t)(N \minus n_\ell), \nonumber\\
 T_{3,\ell}(n_\ell \minus 1|n_\ell) \!&=&\! D n_\ell\left[\frac{N-n_{\ell+1}}{N}+\frac{N-n_{\ell-1}}{N}\right], \nonumber\\
 T_{4,\ell}(n_\ell \plus 1|n_\ell)  \!&=&\! D(N-n_\ell) \left[\frac{n_{\ell+1}}{N}+\frac{n_{\ell-1}}{N}\right], \nonumber \\
 \label{eq:op_transition}
\end{eqnarray}
where the objects ${\ell \!\pm\! 1}$ are to be read as modulo $L$. The constant $D$ regulates how frequently individuals in one site interact with individuals in the neighboring sites. This will ultimately lead to a diffusion-type term in the limiting deterministic description; see below. The quantity $\Pi(n_\ell,t)$ represents a systematic bias towards one of the two states; we will define its functional form below.

The probability distribution, $P(\vec{n},t)$, for the system being in state ${\vec{n} \equals (n_1,\dots,n_L)}$ at time $t$ satisfies the master equation
\begin{equation}
 \frac{d P(\vec{n},t)}{dt}
 \equals \!\sum_{\vecs{n}'\ne\vecs{n}}\!\bigl[T(\vec{n}|\vec{n}') P(\vec{n}',t) \minus T(\vec{n}'|\vec{n}) P(\vec{n},t)\bigr], \label{eq:master}
\end{equation}
where $T(\vec{n}'|\vec{n})$ is the total transition rate from state $\vec{n}$ to state $\vec{n}'$. For convenience, we only indicate the variables that are changed in any one reaction in Eq. (\ref{eq:op_transition}). In the first and third reaction in Eq. (\ref{eq:op_transition}), an up-spin is replaced by a down-spin in site $\ell$. Similarly, the second and fourth reactions describe processes in which a down-spin is replaced by an up-spin. Either process may happen through local interactions within site $\ell$ [first and second reaction in Eq. (\ref{eq:op_transition})], or through interactions with the neighboring sites ${\ell \!\pm\! 1}$ (third and fourth reaction). To interpret the first type of reaction, one may think of an up-spin being chosen for potential update in site $\ell$, hence $T_{1,\ell}$ is proportional to $n_\ell$. This spin is then replaced by a down-spin with rate ${1 \minus \Pi(n_\ell,t)}$. Similarly, the transition rate for the reaction replacing a down-spin by an up-spin through local interaction is proportional to ${N \minus n_\ell}$ and $\Pi(n_\ell,t)$; see the second reaction in Eq. (\ref{eq:op_transition}). 

The third and fourth reactions in Eq. (\ref{eq:op_transition}) finally represent interactions of spins in one site with a neighboring site. In the third reaction, an up-spin in site $\ell$ is chosen for potential update (hence the rate is proportional to $n_\ell$), and then a random spin from either of the two neighboring sites is chosen. If that second spin is in the down-state, the spin in site $\ell$ adopts the down-state as well. The fourth reaction works similarly.

The Ginzburg-Landau potential can be realized as a systematic bias towards the less populous state in the phase of unbroken symmetry, and towards the more populous state in the broken symmetry phase. More specifically, we use
\begin{equation}
 \Pi(n_\ell,t) \equals \frac{1}{2}(1 \plus \phi_\ell) \plus \frac{1}{2}\frac{a}{1\minus2D}\phi_\ell \bigl[g(t) \minus \phi_\ell^2\bigr],\label{eq:op_pi}
\end{equation} 
where $\phi_\ell$ is the local magnetization, ${\phi_\ell \equals (2 n_\ell/N) \minus 1}$ \cite{RB11}.  The parameter ${a \gt 0}$ denotes the strength of the potential; its value needs to be chosen such that ${0 \les \Pi \les 1}$ for all values of ${\phi_\ell \!\in\! [-1,1]}$.  If ${a \equals 0}$, then ${\Pi(n_\ell,t) \!\sim\! n_\ell}$, and there is no bias towards either of the two states. For ${g \lt 0}$, the quantity ${g \minus \phi_\ell^2}$ is negative, and so the second term in Eq. (\ref{eq:op_pi}) describes a bias towards the less populous state. For ${g \gt 0}$ (and assuming ${\phi_\ell^2 \lt g}$),, one has a bias towards the more populous state. The control parameter $g(t)$ can take any real value up to ${g \equals 1}$ in this setup. While Russell and Blythe \cite{RB11} have considered this model at constant values of the parameters, here we will systematically sweep the system across the transition; specifically, we investigate linear quenches of the parameter $g$.

\subsection{Linear-noise approximation}
The master equation describes the above stochastic process exactly, and we recover information about the deterministic dynamics and finite system-size corrections by following the work of van Kampen \cite{vK}. We separate fluctuations, $\xi_\ell$, from the limiting deterministic dynamics and write
\begin{equation}
 \frac{n_\ell}{N} = \frac{1+\phi_\ell^\infty}{2}+N^{-1/2}\xi_\ell, \label{eq:op_fluct}
\end{equation}
where $\phi_\ell^\infty$ represents the local magnetization in site $\ell$ in the deterministic limit. Carrying out the system-size expansion to lowest order, one finds
\begin{equation}
 \dot{\phi}_\ell^\infty = D\Delta\phi_\ell^\infty + a\phi_\ell^\infty\bigl[g(t)-(\phi_\ell^\infty)^2\bigr]. \label{eq:op_meanfield}
\end{equation}
The quantity ${\Delta \phi_\ell^\infty}$ is the lattice Laplacian, i.e., ${\Delta\phi_\ell^\infty \equals \phi_{\ell+1}^\infty \minus 2\phi_\ell^\infty \plus \phi_{\ell-1}^\infty}$. At constant values of $g$, these dynamics have stable spatially homogeneous fixed points,
\begin{equation}
\phi^*(g)=\left\{
  \begin{matrix}
    0 & \mbox{for }\; g \lt 0, \\
    \pm\sqrt{g} & \mbox{for }\; g \gt 0. \\
  \end{matrix}
 \right.
\end{equation}

At next-to-leading order in the expansion of the master equation, we recover a Fokker-Planck equation for the probability distribution of the fluctuations about the mean-field dynamics, $\mathcal{P}(\vec{\xi},t)$, and from this we find the equivalent set of Langevin equations. Anticipating that we will linearize about the zero fixed point of the deterministic dynamics, we simplify these equations using the ${\phi^* \equals 0}$ fixed point of Eq. (\ref{eq:op_meanfield}). This assumption will be justified below. In this linearization, we find that the fluctuations follow the dynamics
\begin{equation}
 \dot{\xi}_\ell = D\Delta\xi_\ell + a g(t)\xi_\ell + \sqrt{\frac{1}{2}}\eta_\ell(t),
\end{equation}
where $\eta_\ell$ is white noise with correlator ${\avg{\eta_\ell(t)\eta_{\ell'}(t')} \equals \delta_{\ell\ell'} \delta(t\minus t')}$. The noise is not correlated across different cells, as there are no reactions which change particle numbers in more than one lattice site at a time.   As we have effectively made the expansion ${\phi_\ell \equals  \phi^* \plus 2N^{-1/2}\xi_\ell}$, and used the fixed point ${\phi^* \equals 0}$, the evolution of the order parameter due to fluctuations about the fixed point is given by
\begin{equation}
 \dot{\phi}_\ell = D\Delta \phi_\ell + a g(t) \phi_\ell + \sqrt{\frac{2}{N}}\eta_\ell(t).\label{eq:op_philinear}
\end{equation}

\subsection{Characteristic length scale and density of defects}
Analytical studies of slow quenches from the disordered into the ordered phase have previously been carried out; see, e.g., \cite{ML99,GM03,DS01,lythe}. For completeness, here we re-iterate the main steps. These analyses start from the Ginzburg-Landau equation (\ref{eq:op_meanfield}), complemented by external additive Gaussian white noise. Taking $g(t)$ to be a linearly increasing function of time, ${g(t) \equals \mu t}$ (${\mu \gt 0}$), and starting at an initial time ${t_0 \lt 0}$, the symmetry-breaking phase transition is crossed at ${t \equals 0}$, hence ${g_b \equals 0}$ and ${g(\hat{t}) \equals g_c}$. Simulations show that the order-parameter field, $\phi$, remains close to zero throughout the stable regime (${g \lt 0}$), and well after the transition point has been crossed. It only ``jumps'' to its non-zero equilibrium value at a well-defined later time, ${\hat{t} \gt 0}$ \cite{lythe,ML99}. This observation provides justification for the linearization about the zero fixed point.

The linearized equation (\ref{eq:op_philinear}) is easily analyzed in Fourier space. We write $\tilde{\phi}_q(t)$ for the Fourier mode of the order parameter with wave number $q$. Thus the structure factor, ${S(q,t) \equals \avg{|\tilde{\phi}_q(t)|^2}}$, where $\avg{\dots}$ represents an average over realizations of the noise, is given by
\begin{equation}
 S(q,t)= \frac{1}{2\pi}\frac{2}{N}e^{a \mu t^2 - 2D q^2 t}\int_{t_0}^{t}dt'\,e^{2D q^2 t' - a\mu t'^2}.
\end{equation}
To evaluate the integral one assumes that $t$ and $t_0$ are sufficiently large for the integral to be well approximated by the infinite limit case [${t_0 \!\to\! -\infty}$, ${t \!\to\! \infty}$; this is justified if $g(t_0)$ and $\hat{g}$ are of order one, and if ${\mu \!\ll\! 1}$]. One makes the further, related assumption that ${D q^2 \!\ll\! 2 a \mu t}$, and the structure factor can be written as
\begin{equation}
 S(q,t) \approx \frac{1}{\sqrt{\pi a \mu}\,N}e^{a \mu t^2 -2D t q^2}. \label{eq:op_sf} 
\end{equation}

Using Parseval's theorem \cite{parseval} the expectation value of $\phi^2(t)$ can be obtained from integrating Eq. (\ref{eq:op_sf}) over $q$. One finds
\begin{equation}
 \avg{\phi^2(t)} \approx \frac{1}{\sqrt{2Da\mu t}\,N}e^{a\mu t^2}. \label{eq:op_phi2}
\end{equation}
To determine when the order parameter jumps from the unstable fixed point to either of the stable fixed points defined by ${\phi^*(t) \equals \pm\sqrt{g(t)}}$, we combine this result with the implicit equation ${\avg{\phi^2(\hat{t})} \equals \delta\hat{g}}$, where ${\hat{g} \equals g(\hat{t})}$ and ${0 \lt \delta \lt 1}$, to find $\hat{g}$ satisfies \cite{ML99}
\begin{equation}
 \hat{g}=\sqrt{\frac{\mu}{a}\ln\left(\delta N \sqrt{2Da}\, \hat{g}^{3/2}\right)}.\label{eq:op_ghat}
\end{equation}
Thus the linear-noise approximation leads to the scaling behavior ${\hat{g} \!\sim\! \mu^{1/2}}$, up to logarithmic corrections, which agrees with the Kibble-Zurek prediction in Eq. (\ref{eq:mech_ghat}).

The structure factor, $S(q,t)$ (${t \gt 0}$), has its peak at ${q \equals 0}$, and accordingly the only length scale in the linearised system is set by its half-width, $\Gamma$, defined by ${S(\Gamma/2,t) \equals S(0,t)/2}$. From Eq. (\ref{eq:op_sf}) we find ${\Gamma(t) \equals \sqrt{2\ln 2/(Dt)}}$, i.e., at the point when the system falls out of the unstable equilibrium near ${\phi \equals 0}$, we have
\begin{equation}
\hat{\Gamma}=\sqrt{\frac{2\mu\ln 2}{D\hat{g}}},\label{eq:op_gammahat}
\end{equation}
with $\hat{g}$ as given above.

The length scale ${\hat{\xi} \!\sim\! \hat{\Gamma}^{-1}}$ is, in an idealised situation, inversely proportional to the number of point defects in the one dimensional system. These defects are identified as zero crossings of the field variable, $\phi$, separating domains of positive and negative order parameter.  The expected density of zero crossings, $\avg{\rho}$, can be estimated using the well-known Liu-Halperin-Mazenko formula \cite{LHM, LHM2}
\begin{equation}
 \avg{\rho(t)} 
  = \frac{1}{\pi}\sqrt{-\frac{\partial_{\ell\ell}\:C(\ell\equals0)}{C(\ell\equals0)}} 
  = \frac{1}{\pi}\sqrt{\frac{\int dq\; q^2 S(q,t)}{\int dq\; S(q,t)}}, \label{eq:numzerosgen}
\end{equation}
where $C(\ell)$ is the spatial equal-time correlation function of the order-parameter field, which is equivalent to the spatial inverse Fourier transform of the structure factor (in order to be able to formally introduce a derivate with respect to $\ell$, a continuation to real $\ell$ is implied). Specifically, for this model one finds
\begin{equation}
 \avg{\rho(\hat{t})} = \frac{1}{2\pi}\frac{1}{\sqrt{D}}\sqrt{\frac{\mu}{\hat{g}}}\sim \mu^{1/4}.\label{eq:op_numzeros}
\end{equation}
These results reproduce those of Ref. \cite{ML99}, and the resulting scaling of the density of defects with the quench rate is in agreement with Eq. (\ref{eq:mech_zeros}). The density of zero-crossings is related to the width of the structure factor through the relation
\begin{equation}
 \avg{\rho(t)} = \frac{1}{2\pi}\frac{1}{\sqrt{2\ln2}} \Gamma(t). \label{eq:numzeros_gamma}
\end{equation}

\subsection{Test against simulations}
\begin{figure}[t]\includegraphics[width=1.\columnwidth]{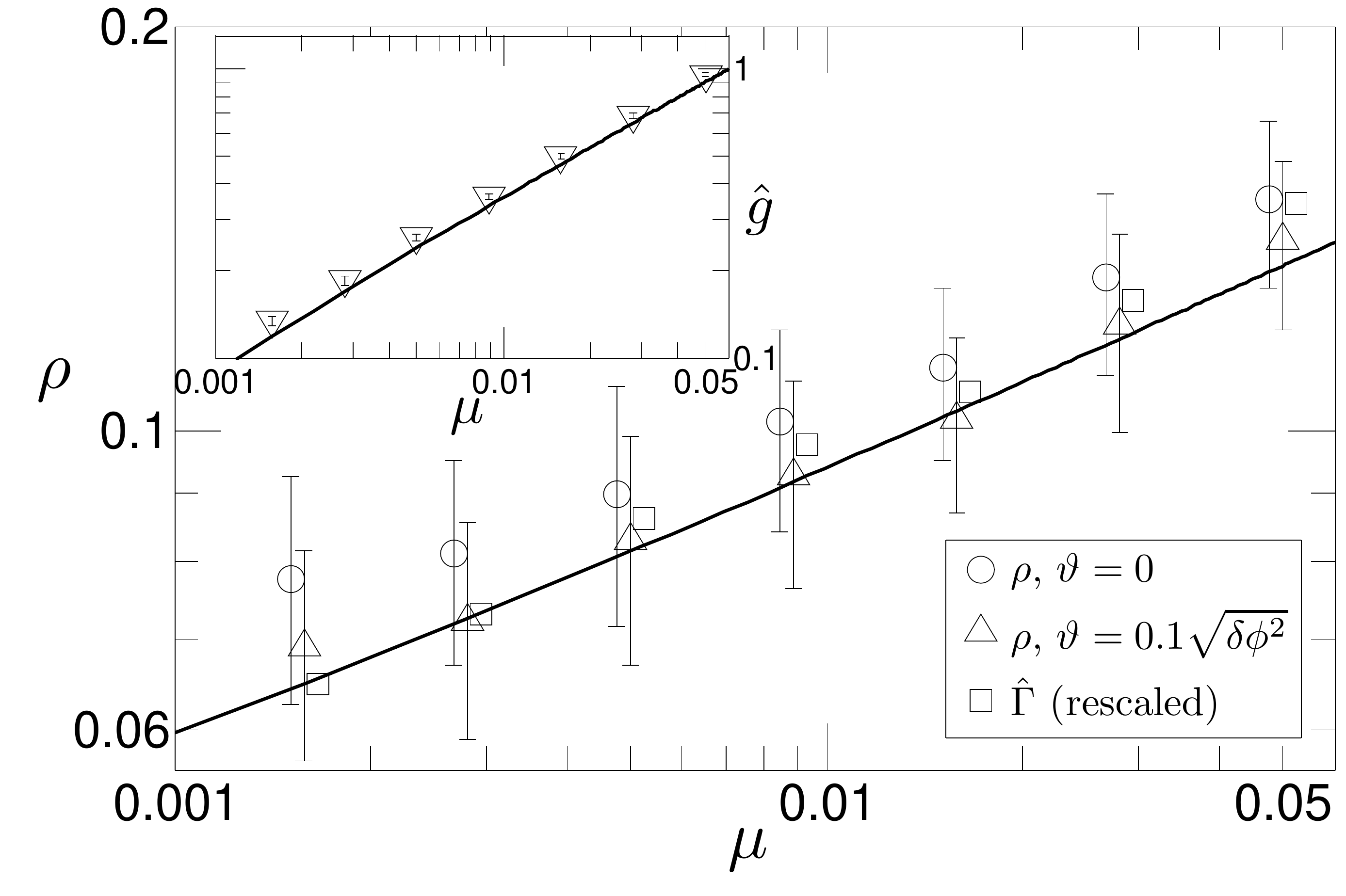}
\caption{Density of defects per unit length in the model of opinion dynamics at ${g \equals \hat{g}}$. Defects are counted directly (circles) and with an imposed threshold (triangles); see text for details. Data are from stochastic simulations of the opinion dynamics model [Eq. (\ref{eq:op_transition})]. Also shown (squares) is the width of the structure factor, $\hat{\Gamma}$, from the stochastic simulations,  re-scaled by a factor of $2\pi\sqrt{2\ln 2}$ to collapse with $\rho$ [see Eq. (\ref{eq:numzeros_gamma})]. The solid line is the theoretical prediction of Eq. (\ref{eq:op_numzeros}). Inset: Values of $\hat{g}$ from stochastic simulations (symbols) and from the theory, Eq. (\ref{eq:op_ghat}) (solid line). Error bars represent standard deviations over $100$ realizations. Model parameters are ${a \equals 0.42}$, ${D \equals 0.08}$, ${N \equals 2 \timess 10^4}$, ${L \equals 200}$, and ${\delta \equals 0.2}$. }
\label{fig:fig1}
\end{figure}

We compare quantitative predictions of the linear-noise approximation, Eqs. (\ref{eq:op_ghat}), (\ref{eq:op_gammahat}) and (\ref{eq:op_numzeros}) against numerical simulations in Fig. \ref{fig:fig1}. Simulations of the stochastic process described by reactions (\ref{eq:op_transition}) are performed here using the stochastic simulation algorithm by Gillespie \cite{gillespie,remark:gillespie}. The range of the quench rate, $\mu$, over which we can obtain results has a lower limit due to difficulties in counting zero crossings when $\hat{g}$ is close to zero, and an upper limit as the microscopic model is only meaningful for ${g(t) \les 1}$. As seen in the inset, theoretical predictions for $\hat{g}$ agree with simulations. Similarly, good agreement is found for the width of the structure factor, $\hat{\Gamma}$, at the point at which the defects are formed.
The width is measured by fitting a Gaussian to the averaged structure factors calculated from the simulations at ${t \equals \hat{t}}$. We find that the fitting process is susceptible to errors when the structure factor is not sharply peaked. To avoid this, we choose the diffusion rate, $D$, to be sufficiently large.
Counting the number of defects (i.e., zerocrossings) directly comes with some difficulty not previously reported for other systems \cite{ML99,GM03}. Those existing studies have mostly focused on very small amplitudes of external noise, typically of the order $10^{-8}$ or so (see, e.g., \cite{ML99}). In our model, the source of the noise is not external, but intrinsic, and its amplitude is proportional to $N^{-1/2}$, where $N$ is the number of individuals in each site. In the simulations leading to Fig. ~\ref{fig:fig1} we use ${N \equals 2 \timess 10^4}$, which corresponds to a noise amplitude several orders of magnitude above those typically used for direct simulations of defect formation in stochastic partial differential equations. We find that a naive counting of zero crossings gives results consistently above the predictions from the theory, see Fig. \ref{fig:fig1}, and deviations are particularly high at small quench rates when the freeze-out occurs close to ${g \equals 0}$. We attribute this to the relatively large noise amplitude, leading to spurious zero crossings in the simulations, and to the fact that defects may not have fully formed. Further analysis has shown that the agreement of stochastic simulations with the analytic solutions improves with increasing system size, $N$. To avoid counting spurious zeros, we empirically introduce a threshold to ensure the kinks satisfy a minimum size requirement. Specifically, we only count two zero crossings of the order-parameter field as separate defects provided the magnitude of the order-parameter field exceeds a threshold, $\vartheta$, in between. This threshold is chosen to be a fraction of the rms field amplitude, which at time $\hat{t}$ is given by ${\sqrt{\delta [\phi^*(\hat{t})]^2}}$, where ${\phi^*(\hat{t})\equals\pm\sqrt{\hat{g}}}$ represents the stable fixed point of Eq. (\ref{eq:op_meanfield}). Specifically, we choose ${\vartheta \equals 0.1 \timess \sqrt{\delta [\phi^*(\hat{t})]^2}}$, and applying this procedure we find that simulation results are close to the predictions of the linear theory; see the main panel of Fig. \ref{fig:fig1}.   

\subsection{Discussion}

Two remarks are in order before we move to a more complex example in the next section. First, the analytical results presented in the previous section are, up to constants and normalization factors, identical to those obtained previously in Ref. \cite{ML99}. The main difference is the source of noise in the model. In most existing studies, noise was added externally to a deterministic partial differential equation. In our example, we start from an individual-based model, in which the noise is intrinsic and originates from the stochastic reaction dynamics in finite populations. Carrying out the system-size expansion, we ultimately arrive at an equation very similar to those studied previously (the Ginzburg-Landau equation); the microscopic model was designed to do so. The fact that the noise comes out as white noise in the Gaussian approximation is again a feature of the specific microscopic model we used as a starting point. We chose this simple example to make contact with existing studies of defect formation in slow quenches. We will move to more complex models below. 
The second remark concerns the application of the threshold, $\vartheta$, to identify the relevant defects. This is an ad hoc procedure; a detailed analysis shows that the absolute number of defects counted carries some dependence on the threshold. We choose the threshold to be one-tenth of the root-mean-square (rms) field amplitude at the point at which the defects are counted, and we find that this leads to good agreement with the theoretical predictions, although we do not have further justification for this choice. It is important to keep in mind that the Liu-Mazenko-Halperin formula, Eq. (\ref{eq:numzerosgen}), is subject to various constraints, in particular a continuous-order parameter field, and that similar problems relating to spurious zeros are briefly mentioned, for example, in Refs. \cite{LytheThesis,YZ98}. We would argue that the length scale set by the width of the structure factor is the more fundamental quantity here, and that the density of defects has more of a derived character. For the width of the structure factor, we observe very good agreement between theory and simulations, and in this sense we think that the KZ theory is perfectly applicable to the individual-based system we study here.

\section{Evolutionary dynamics}\label{sec:evol}
\subsection{Model definition}
We next consider an example from the theory of evolutionary dynamics, more specifically a spatial meta-population model of two species who interact subject to natural selection and mutation.  As in Sec. \ref{sec:opinion}, the model is defined on a one-dimensional periodic lattice with $L$ sites, and it operates in continuous time. Each site hosts a well-mixed population of individuals, each of which can be of type $A$ or of type $B$. These represent the two interacting species or phenotypes. We write $n_\ell$ for the number of individuals of type $A$, and $m_\ell$ for the number of individuals of type $B$ in site $\ell$. The interaction between the two phenotypes is governed by an evolutionary game, defined by the so-called payoff matrix
 \begin{equation}
 \begin{array}{c|cc}
  & A & B \\ \hline
  A & 1 & 1\minus g(t) \\
  B & 1\minus g(t) & 1
 \end{array}.
\end{equation}
Details of stochastic evolutionary game theory can be found, for example, in \cite{hauert}. Broadly speaking, an interaction between two individuals of the same type ($AA$ or $BB$) adds one unit of fitness to each of their reproductive propensities, while an encounter of individuals of two different types (an $A$ and a $B$) contributes ${1 \minus g}$ to each of their fitnesses. We use $g$ as a time-dependent external control parameter; its interpretation will be discussed in more detail below. 

The expected (re-scaled) fitness of an individual of type $A$ (respectively $B$) in site $\ell$ is then given by
\begin{eqnarray}
 \Pi_A(n_\ell,m_\ell,t)&=&\frac{n_\ell}{\Omega} + \frac{m_\ell}{\Omega}\bigl[1\minus g(t)\bigr],\nonumber\\
 \Pi_B(n_\ell,m_\ell,t)&=&\frac{n_\ell}{\Omega}\bigl[1\minus g(t)\bigr] + \frac{m_\ell}{\Omega}.
\end{eqnarray}
In our model, the total number of individuals in a given site, ${n_\ell \plus m_\ell}$, will not be constant, and so we have introduced $\Omega$ as the {\em typical} number of individuals in each lattice site. We will use $\Omega^{-1/2}$ as the expansion parameter.

Reactions between individuals within a given site occur with the following transition rates:
\small
\begin{eqnarray}
 T_{B_\ell\to A_\ell}(n_\ell,m_\ell) &=& \frac{1}{2}(1-\nu)\left[1+\beta\bigl(\Pi_A-\Pi_B\bigr)\right]\frac{n_\ell m_\ell}{\Omega} \nonumber\\
 &&+\frac{1}{2}\frac{\nu}{\Omega}m_\ell^2, \nonumber \\ 
 T_{A_\ell\to B_\ell}(n_\ell,m_\ell) &=& \frac{1}{2}(1-\nu)\left[1+\beta\bigl(\Pi_B-\Pi_A\bigr)\right]\frac{n_\ell m_\ell}{\Omega} \nonumber\\
 &&+\frac{1}{2}\frac{\nu}{\Omega}n_\ell^2. \label{eq:evo_transitionevo}
\end{eqnarray}
\normalsize
The first reaction describes transitions in which an individual of type $B$ is converted into an individual of type $A$, and the second reaction describes the opposite process. A conversion of, say, a $B$ into an $A$ can occur via two different routes: (i) two individuals of different types interact, and conversion of the $B$ into an $A$ occurs with a rate proportional to ${[1 \plus \beta(\Pi_A \minus \Pi_B)]/2}$. The opposite conversion happens with rate ${[1 \plus \beta(\Pi_B \minus \Pi_A)]/2}$. This is known as the ``pairwise local comparison process'' \cite{TCH05}; the parameter $\beta$ indicates the strength of selection. For ${\beta \equals 0}$, the relative fitnesses of the two types of individuals are irrelevant, and the dynamics describes neutral evolution. For ${\beta \gt 0}$, differences in fitness increasingly matter. The parameter $\nu$ is a mutation rate, indicating the rate with which copying errors occur. Thus in an interaction of an $A$ and a $B$, in which one is chosen for reproduction and the other for removal, an effective change of $n_\ell$ and $m_\ell$ only occurs when no copying error is made, i.e., with a rate proportional to ${1 \minus \nu}$. These processes are described by the first term in each of the transition rates given above. (ii) As a consequence of copying errors, effective changes of $n_\ell$ and $m_\ell$ may result from an interaction of two individuals of the same type. This occurs with a rate proportional to $\nu$ and is captured by the second term in each of the above reaction rates. 
In addition to the on-site reactions, we allow particle hopping between neighboring sites with the following rates:
 \begin{eqnarray}
 T_{A_{\ell}\to A_{\ell'}}(n_\ell)  &=& D n_\ell \, \delta_{|\ell-\ell'|,1}, \nonumber\\
 T_{B_{\ell}\to B_{\ell'}}(m_\ell) &=& D m_\ell \, \delta_{|\ell-\ell'|,1}.
  \label{eq:evo_transitiondiff}
\end{eqnarray}
The first of these two reactions describes the hopping of a particle of type $A$ from site $\ell$ to a neighboring site ${\ell' \equals \ell \!\pm\! 1}$, while the second reaction captures the hopping of individuals of type $B$. The parameter ${D\gt0}$ represents the hopping rate. We stress that this is not an exchange process, but that the total particle numbers in each of the two cells change.

\subsection{Linear-noise approximation and characteristic length scale}
To carry out the system-size expansion, we write ${(\vec{\psi}^{\bf \infty},\vec{\chi}^{\bf \infty})}$ for the deterministic concentrations of individuals of type $A$ and $B$, respectively, i.e., ${\psi_\ell^\infty \equals \lim_{\Omega\to\infty} n_\ell/\Omega}$, and similarly for $\chi_\ell^\infty$. To capture Gaussian fluctuations, we write
\begin{equation}
 \frac{n_\ell}{\Omega} \mapsto \psi_\ell^\infty + \Omega^{-1/2}\xi_\ell, \quad
 \frac{m_\ell}{\Omega} \mapsto \chi_\ell^\infty + \Omega^{-1/2}\zeta_\ell,
\end{equation}
where ${(\vec{\xi},\vec{\zeta})}$ are the variables which represent the stochastic contributions to the dynamics, within the first sub-leading order of the van Kampen expansion.

From the leading-order terms in the expansion, we recover a system of equations which describes the evolution of the deterministic concentrations,
\small
\begin{subequations}
\begin{eqnarray}
 \dot{\psi}_\ell^\infty
  \!&=&\! D\Delta\psi_\ell^\infty 
  + (1-\nu)\beta g(t) \psi_\ell^\infty \chi_\ell^\infty (\psi_\ell^\infty \minus \chi_\ell^\infty )\nonumber\\
  &&\qquad\qquad
  -\frac{\nu}{2}\bigl[(\psi_\ell^\infty)^2 \minus( \chi_\ell^\infty)^2\bigr], \label{eq:evo_psi} \\
 \dot{\chi}_\ell^\infty
  \!&=&\! D\Delta\chi_\ell^\infty 
  - (1-\nu)\beta g(t) \psi_\ell^\infty\chi_\ell^\infty(\psi_\ell^\infty \minus \chi_\ell^\infty)\nonumber\\
  &&\qquad\quad
  +\frac{\nu}{2}\bigl[(\psi_\ell^\infty)^2 \minus (\chi_\ell^\infty)^2\bigr]. \label{eq:evo_chi}
\end{eqnarray}
\end{subequations}
\normalsize
As before, $\Delta$ is the discrete Laplacian operator. It is convenient to introduce the order-parameter field ${\phi_\ell \equals (n_\ell \minus m_\ell)/\Omega}$, which in the deterministic limit, written as $\phi_\ell^\infty$, is simply the difference of the concentrations, $\psi_\ell^\infty$ and $\chi_\ell^\infty$. In the finite system, the total number of particles in each site is of order $\Omega$, and we expect site-to-site fluctuations to be of order $\Omega^{1/2}$. In the deterministic limit, these fluctuations become irrelevant. Assuming from now on that initial conditions are such that ${n_\ell \plus m_\ell \equals \Omega}$ for all $\ell$, we have ${\psi_\ell^\infty \plus \chi_\ell^\infty \equals 1}$ in the deterministic limit. The evolution of the order parameter is then described by
\begin{equation}
 \dot{\phi}_\ell^\infty = D\Delta\phi_\ell^\infty 
  + \frac{1-\nu}{2}\beta g(t)\bigl[1-(\phi_\ell^\infty)^2\bigr]\phi_\ell^\infty
  -\nu\phi_\ell^\infty. \label{eq:evo_phi}
\end{equation}

For constant values of $g$, the stable fixed points of these dynamics are
 \begin{equation}
 \phi^*(g)=
 \left\{
  \begin{matrix}
    0 & \mbox{for }\; g \lt \frac{2\nu}{(1-\nu)\beta}, \\
    \pm\sqrt{1-\frac{2\nu}{(1-\nu)\beta g}} & \mbox{for }\; g \gt \frac{2\nu}{(1-\nu)\beta}, \\
  \end{matrix}
 \right.\label{eq:evo_fp}
\end{equation}
and the bifurcation point is ${g_b \equals 2\nu/[(1\minus\nu)\beta]}$.

By linearizing the next-to-leading-order equation about the ${\phi^* \equals 0}$ fixed point, one obtains the Langevin equation 
\begin{equation}
\dot{\lambda}_\ell=\left[D\Delta + \frac{1-\nu}{2}\beta g(t)-\nu\right]\lambda_\ell+\eta_\ell(t) \label{eq:evo_lambda}
\end{equation}
for the quantity ${\lambda_\ell \equals \xi_\ell \minus \zeta_\ell}$, which represents the fluctuations about the fixed point. The main difference between the analysis here and that in Sec. \ref{sec:opinion} is that the hopping reactions result in spatially correlated noise terms, $\eta_\ell$. Specifically, we have
\begin{equation}
\avg{\eta_\ell(t)\eta_{\ell'}(t')}=\delta(t-t')\left[(1+4D)\delta_{\ell,\ell'}-2D\delta_{|\ell-\ell'|,1}\right]. \label{eq:evo_noise}
\end{equation}

Further details of the derivation can be found in Appendix \ref{app:noise}. Switching again to Fourier space, the structure factor, ${S(q,t) \equals \avg{|\tilde{\lambda}_q(t)|^2}}$, takes the form
\small
\begin{eqnarray}
 S(q,t) \!&=&\! \frac{1+4D(1-\cos q)}{2\pi}\; e^{\frac{1-\nu}{2}\beta \mu t^2 - 2(\nu+Dq^2)t} \times \nonumber\\
  && \qquad \int_{t_0}^{t} dt'\; e^{-\frac{1-\nu}{2}\beta \mu t'^2 + 2(\nu+Dq^2)t'}.
\end{eqnarray}
\normalsize

As in Sec. \ref{sec:opinion}, one assumes that $t$ and $t_0$ are sufficiently large for the integral to be well approximated by the infinite limit case. Furthermore, the structure factor is sharply peaked about ${q \equals 0}$, so one assumes ${(\nu \plus Dq^2) \!\ll\! (1 \minus \nu)\beta \mu t}$ and ${1 \minus \cos q \!\approx\! q^2/2}$. Making these approximations, the structure factor can be written as
\begin{equation}
 S(q,t) \approx \frac{1+2Dq^2}{\sqrt{2\pi(1-\nu)\beta\mu}}e^{\frac{1-\nu}{2}\beta\mu t^2-2(\nu+Dq^2)t}, 
\end{equation}
and integrating over $q$, one finds the expectation value of $\phi^2(t)$,
\begin{equation}
 \avg{\phi^2(t)} = \frac{1+\frac{1}{2t}}{2\sqrt{(1-\nu)D\beta\mu t}\,\Omega}\;e^{\frac{1-\nu}{2}\beta\mu t^2-2\nu t}. \label{eq:evo_phi2}
\end{equation}

The time, $t_c$, at which the order parameter jumps from the unstable fixed point, ${\phi^* \equals 0}$, to the stable point can be found from
\begin{equation}
 \avg{\phi^2(t_c)} = \delta \left[1 - \frac{2\nu}{(1-\nu)\beta g(t_c)}\right], \label{eq:evo_ghat}
\end{equation} 
where the term in square brackets on the right-hand side is the square of the stable fixed point in the ordered phase, as given in Eq. (\ref{eq:evo_fp}). To estimate the expected number of zero crossings, finally, we again use the Liu-Mazenko-Halperin formula and find
\begin{equation}
 \avg{\rho(t_c)} = \frac{1}{2\pi} \frac{1}{\sqrt{D}}\sqrt{\frac{\mu}{g_c}}\sqrt{1+\frac{\mu}{\frac{\mu}{2}+g_c}}, \label{eq:evo_numzeros}
\end{equation}
where ${g_c \equals g(t_c)}$.

\subsection{Test against simulation}
\begin{figure}[t]\includegraphics[width=\columnwidth]{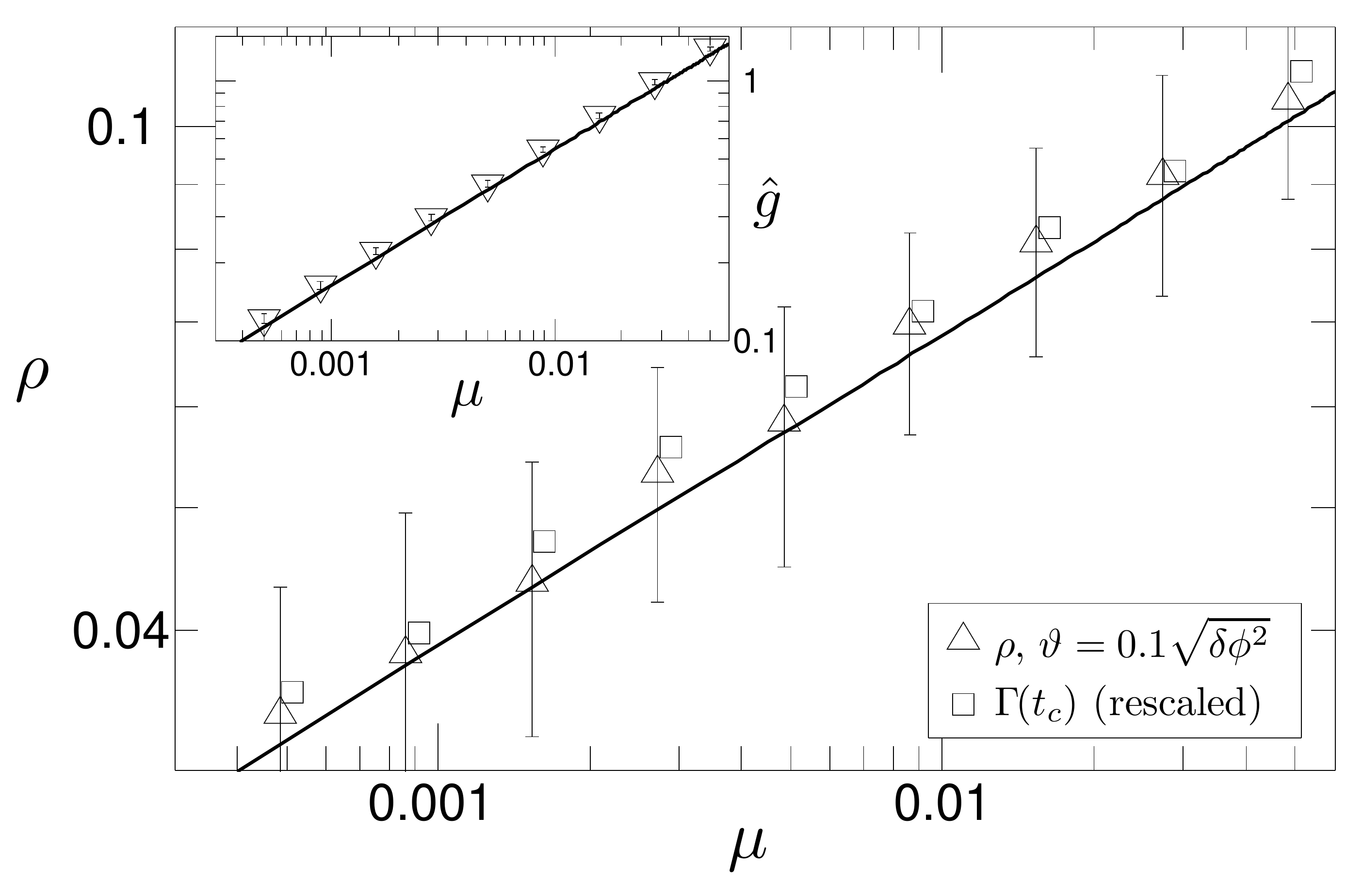}
\caption{Density of defects in the model of evolutionary dynamics at ${g \equals g_c}$. Triangles are from simulations of the individual-based model, Eqs. (\ref{eq:evo_transitionevo}) and (\ref{eq:evo_transitiondiff}). As before, a threshold has been applied when counting zero crossings (see the text). Squares represent simulation data for the width of the structure factor, $\Gamma(t_c)$, rescaled to collapse with $\rho$. The solid line is from Eq. (\ref{eq:evo_numzeros}). Inset: Symbols show measurements of $\hat{g}$ from simulations. The solid line shows ${\hat{g} \equals g(t_c) \minus g_b}$, with $t_c$ as obtained from Eq. (\ref{eq:evo_ghat}). Error bars represent standard deviations over $100$ realizations. Model parameters are ${\nu \equals 0.001}$, ${\beta \equals 0.38}$, ${D \equals 0.1}$, ${\Omega \equals 5000}$, ${L \equals 200}$, and ${\delta \equals 0.2}$.}
\label{fig:fig2}
\end{figure}

\begin{figure}[t]\includegraphics[width=\columnwidth]{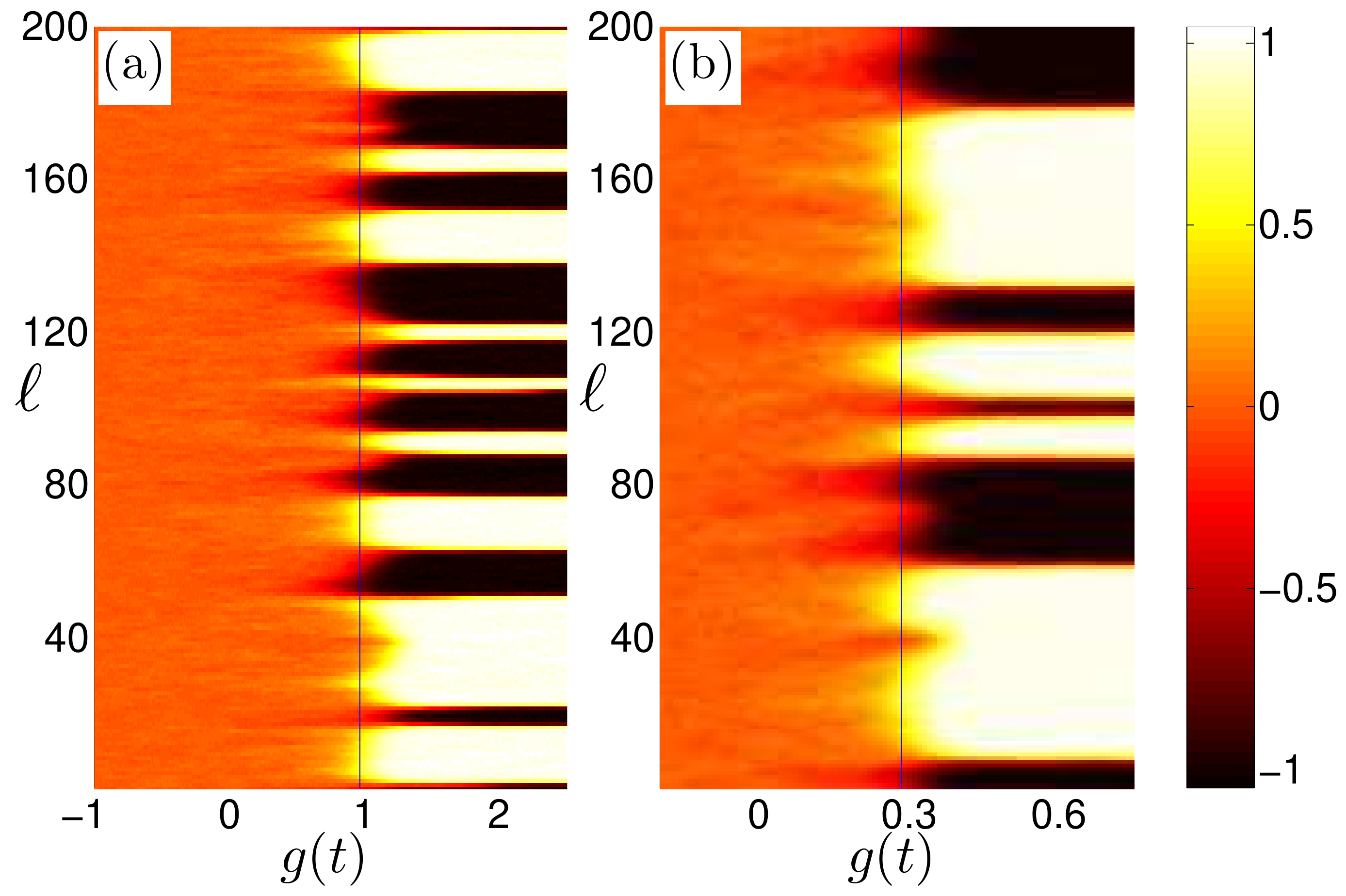}
\caption{(Color on-line). Spatio-temporal dynamics of the order parameter, $\phi_\ell(t)$, for the model of evolutionary dynamics. Light shading indicates high values of $\phi_\ell(t)$, dark shading indicates low values. Solid line corresponds to the value of $g_c$ for this realization of the stochastic dynamics. The quench rates are: (a) ${\mu \equals 2.8 \timess 10^{-2}}$; (b) ${\mu \equals 2.8 \timess 10^{-3}}$. The remaining model parameters are as in Fig. \ref{fig:fig2}.}
\label{fig:fig3}
\end{figure}

We test these analytical predictions against simulations in Fig. \ref{fig:fig2}. As in the previous section, simulations of the microscopic model, Eqs. (\ref{eq:evo_transitionevo}) and (\ref{eq:evo_transitiondiff}), are carried out using the Gillespie algorithm. We vary the quench rate, $\mu$, from ${5 \timess 10^{-4}}$ to ${5 \timess 10^{-2}}$, which for our parameters satisfies the requirement that the transition rates must be positive \cite{remark:beta}.
As in the previous model, we find excellent agreement between simulations and theory for the quantity ${\hat{g} \equals g(t_c) \minus g_b}$, representing the amount of delay experienced by the bifurcation; see the inset of Fig. \ref{fig:fig2}. The density of defects, i.e., zero crossings of the order-parameter field, $\phi(t_c)$, is subject to the same difficulties as in the previous model, and so we again apply an empirical threshold $\vartheta$ to eliminate spurious defects. As before, we choose ${\vartheta \equals 0.1 \timess \sqrt{\delta[\phi^*(t_c)]^2}}$, where $\phi^*(t_c)$ is the stable fixed point in the ordered phase, now given by Eq. (\ref{eq:evo_fp}). As seen in the main panel of Fig. \ref{fig:fig2}, this leads to very good agreement with the theoretical predictions [Eq. (\ref{eq:evo_numzeros})]. 
We also see excellent agreement for the measured width of the structure factor, $\Gamma(t_c)$, from simulations. As before, $\Gamma(t_c)$ is rescaled in Fig. \ref{fig:fig2} for optical convenience, using Eq. (\ref{eq:numzeros_gamma}), in order to agree with $\avg{\rho}$. Examples of the patterns formed by the system are shown in Fig. \ref{fig:fig3} for two different values of the quench rate $\mu$. The light and dark shaded regions represent populations dominated by individuals of types $A$ and $B$, respectively. As seen in the figure, fast quenches result in multiple domains, each of a relatively small size [Fig. \ref{fig:fig3}(a)], whereas slow quenches produce relatively few large-scale domains [Fig. \ref{fig:fig3}(b)].

While the symmetry breaking leading to the formation of the domains in Fig. \ref{fig:fig3} is triggered by intrinsic noise, the origin of the patterns is qualitatively different from the mechanism underlying stochastic Turing patterns. The wavelength of the patterns shown in Fig. \ref{fig:fig3} is set by the quench rate, $\mu$, and the amplitude is not proportional to the noise intensity.
To distinguish the two phenomena, it is also useful to realize that stochastic Turing patterns are sustained by noise (i.e., switching the noise off once the patterns have emerged willl remove them), whereas the patterns shown in Fig. \ref{fig:fig3} are triggered by noise, but will remain if the noise is switched off once the domains have formed.

\section{Decision making of cells}\label{sec:cells}
\subsection{Deterministic model}
The dynamical process of a system being swept slowly across a symmetry-breaking bifurcation has an interesting application in the modeling of cell differentiation. In his now famous picture of an ``epigenetic landscape,'' Waddington represents a cell by a ball or marble rolling down a landscape of bifurcating valleys \cite{waddington}. As time progresses and the marble rolls downhill these valleys may split, and the cell (or marble) has to make a decision about which path to take. In Waddington's metaphorical picture, this represents cell differentiation. These ideas have been applied to gene regulatory systems in a number of biological systems. Most notable are the so-called toggle switches, for example, in the context of the development of drosophila embryos \cite{JIM12} or E. coli \cite{GCC00}. One common class of simple models comprises two fate-determining biological agents \cite{Ferrell12} (for example, transcription factors) with mutually inhibitory interaction. A simple deterministic model capturing the salient features is given by the following set of differential equations:
\begin{eqnarray}
 \dot{\psi} &=&   \frac{1}{1 \plus \bigl(g\chi\bigr)^\gamma} - \beta\psi,  \nonumber \\
 \dot{\chi} &=&  \frac{1}{1 \plus \bigl(g\psi\bigr)^\gamma} - \beta\chi.\label{eq:bio_psichi}
\end{eqnarray}
The variables $\psi$ and $\chi$ describe the concentrations of the two competing substances. The first term in each reaction describes mutual inhibition; the growth rate of either substance is suppressed by the presence of the other reactant.
The variable $g$ controls the strength of this interaction. The terms proportional to $\beta$ finally are decay terms. While the suppression terms follow the commonly used Hill functional form \cite{hill, Ferrell12}, this is of course a rather stylized model. Decay rates, interaction coefficients, and Hill coefficients could in principle differ among the substances, and other reactants have been neglected. Our aim is not to construct a detailed model of any particular biological system, but instead to study the main principles at work. The non-spatial model defined above displays the required bi-stability. More precisely, these equations have a symmetric fixed point, ${\psi^* \equals \chi^*}$, given by
\begin{equation}
 \chi^*(g)=\psi^*(g)=\frac{1}{\beta}\frac{1}{1 \plus \bigl(g\psi^*\bigr)^\gamma}, \label{eq:bio_fixed}
\end{equation}
where we note the dependence on $g$. This fixed point is stable for ${g \lt g_b}$, and unstable for ${g \gt g_b}$, where $g_b$ is the bifurcation point, which can be found from linear stability analysis. For ${g \gt g_b}$, two additional fixed points are found. These are stable and have to be calculated numerically. The resulting phase portraits in the two phases are illustrated in Fig. \ref{fig:fig4}. For coupling strengths smaller than a critical value, ${g \lt g_b}$, the system has a unique fixed point at relatively high concentrations of both reactants, as shown in Fig.~{\ref{fig:fig4}(a). This corresponds to the undifferentiated state, a unique valley in Waddington's landscape picture. For ${g \gt g_b}$, however, the symmetric fixed point is unstable, and two stable attractors emerge, as shown in Fig.~\ref{fig:fig4}(b). At each of these stable attractors, one substance dominates over the other, corresponding to a differentiated state. Throughout this section, we use the parameter values ${\beta \equals 0.5}$ and ${\gamma \equals 4}$, so that the symmetry-breaking bifurcation occurs at ${g_b \equals 2 \timess 3^{-5/4} \!\approx\! 0.5}$.

\begin{figure}[t]\includegraphics[width=\columnwidth]{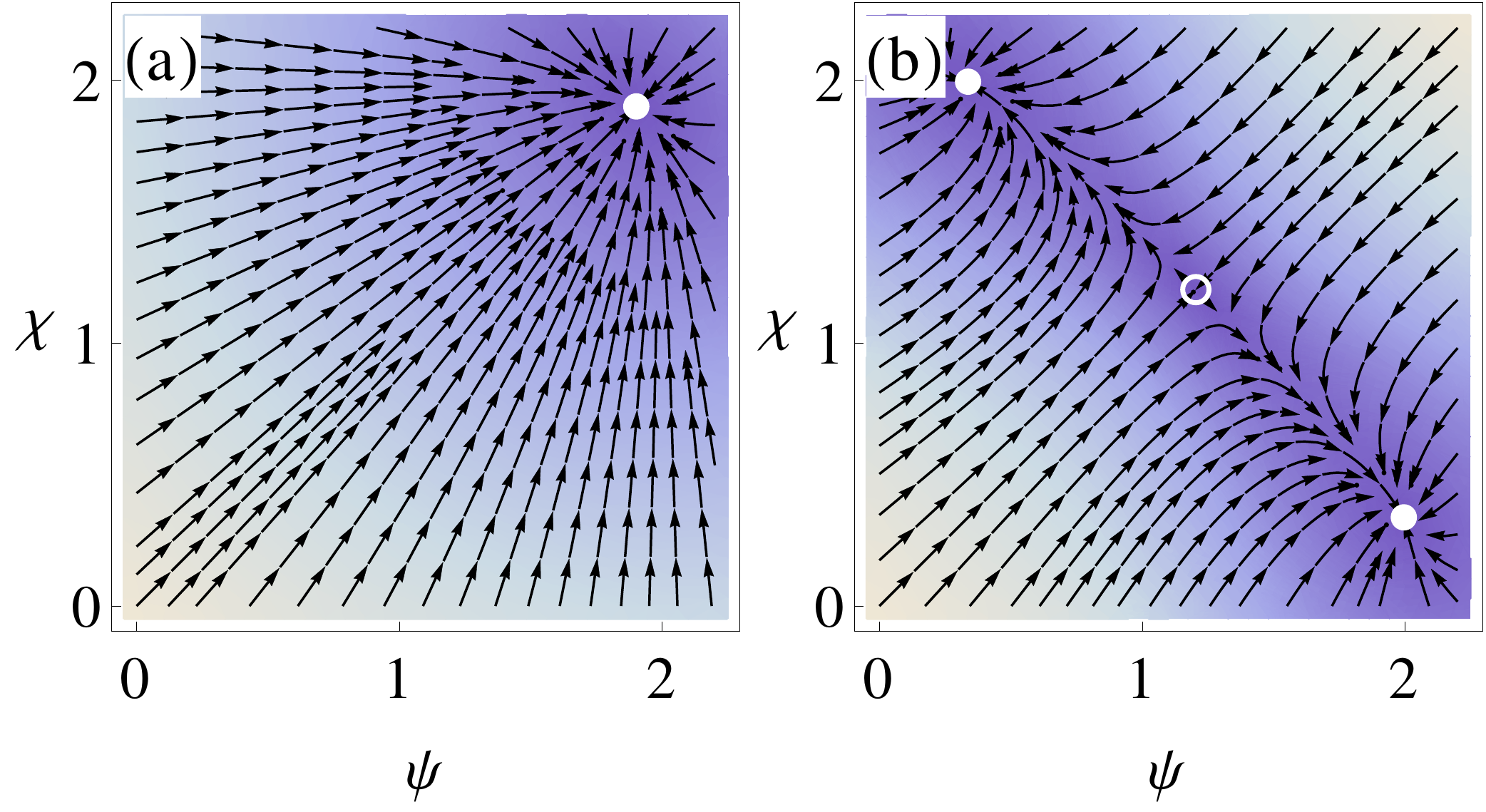}
\caption{(Color on-line) Phase portraits of the deterministic model of cell differentiation, Eq. (\ref{eq:bio_psichi}), for ${g \equals 0.25}$ (a) and ${g \equals 0.75}$ (b). Arrows indicate the flow of the dynamics, filled circles are stable fixed points, and the open circle indicates an unstable fixed point. The background color indicates the speed of the flow, ${\sqrt{\dot{\psi}^2+\dot{\chi}^2}}$.}
\label{fig:fig4}
\end{figure}

In this section we will consider an individual-based spatial realization of this model, subject to a continuous sweep of $g$ from the undifferentiated regime to the differentiated phase. As in the previous sections, $g$ is swept linearly in time, such that ${g(t) \equals \mu t}$. As we will see, the KZ picture is readily applicable to this scenario, and good predictions can be made about the spatial patterning resulting from such a protocol. 
 
To illustrate the decision-making process, we show the evolution of an individual-based realization of the above model in Fig. \ref{fig:fig5}. The exact model will be defined further below. The continuous smooth lines indicate the location of the stable fixed points of the system as $g$ is varied, dashed lines represent unstable fixed points. The fluctuating lines are the particle concentrations, $\psi$ and $\chi$. As seen in the figure these stay close to the symmetric fixed point in the initial phase of the evolution, ${g \lt g_b}$, but also into the symmetry-broken phase, ${g \gt g_b}$, when the symmetric fixed point is unstable. Symmetry breaking only occurs some time into the broken phase at a time $t_c$, when ${g(t_c) \gt g_b}$. This delay in the bifurcation depends on the quench rate, $\mu$. As seen in Fig. \ref{fig:fig5}(a), the delay can be significant for fast quenches, but it is reduced in slower quenches [Fig. \ref{fig:fig5}(b)]. At this freeze-out, the system makes its ``decision,'', and one of the concentrations, $\psi$ or $\chi$, will assume a relatively low value while the other one will assume a significantly higher value, corresponding to the two stable fixed points of the system in the symmetry-broken phase.
  
\begin{figure}[t]\includegraphics[width=\columnwidth]{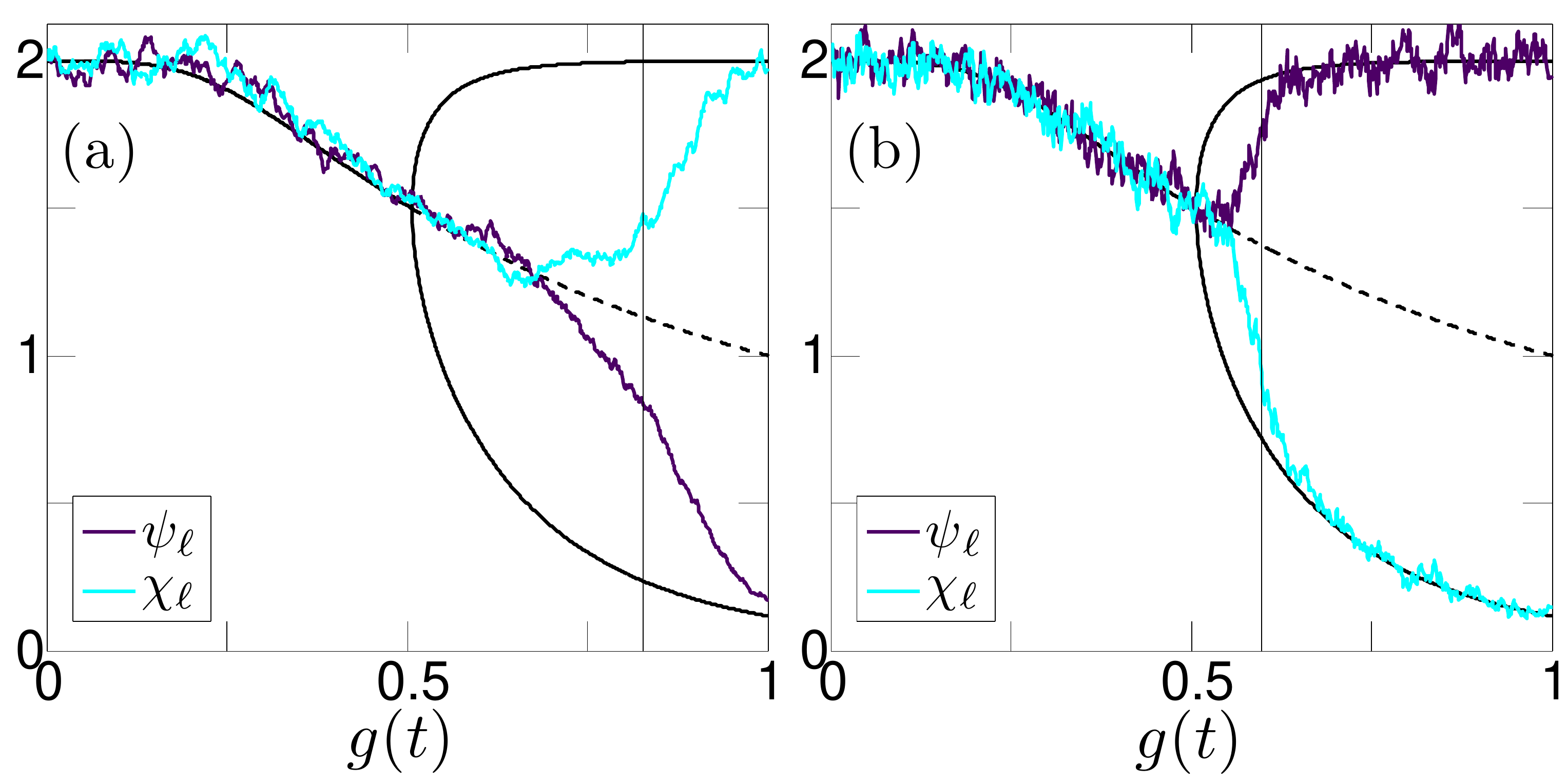}
\caption{ (Colour on-line). Trajectories obtained from single simulation runs of the stochastic model of cellular decision making. The noisy lines show simulation data for $\psi_\ell$ and $\chi_\ell$ at a single lattice site. The quench rates are: (a) ${\mu \equals 2.3 \timess 10^{-2}}$; (b) ${\mu \equals 2.3 \timess 10^{-3}}$. The solid smooth lines indicate stable fixed points of the deterministic dynamics; dashed lines are unstable symmetric fixed points. Model parameters are ${D \equals 0.1}$, ${V \equals 1000}$, ${L \equals 200}$, and ${\delta \equals 0.2}$.}
\label{fig:fig5}
\end{figure}

\subsection{Definition of the individual-based model}
The model is again defined on a one-dimensional periodic lattice with $L$ sites, with each lattice site $\ell$ containing $n_\ell$ molecules of the first chemical reactant, and $m_\ell$ of the second. Each lattice site represents a biological cell in this setup, and we denote the cell volume by $V$. In the deterministic limit, the concentrations of the two chemicals in cell $\ell$ are given by ${\psi_\ell^\infty \equals \lim_{V\to\infty} (n_\ell/V)}$ and ${\chi_\ell^\infty \equals \lim_{V\to\infty} (m_\ell/V)}$. The production of molecules of either type occurs with rates
\begin{eqnarray}
 T_{1,\ell}(n_\ell \plus 1,m_\ell|n_\ell,m_\ell) &=&  h(m_\ell/V,t)\,V, \nonumber\\
 T_{2,\ell}(n_\ell,m_\ell \plus 1|n_\ell,m_\ell) &=&  h(n_\ell/V,t)\,V, \label{eq:bio_transition1}
\end{eqnarray}
where $h(x,t)$ is the inhibitory Hill function,
\begin{equation}
 h(x,t) = \frac{1}{1 \plus [g(t)x]^\gamma},\label{eq:bio_hill}
\end{equation}
see also \cite{JIM12}. 
Both substances decay with rate $\beta$; these reactions are described by
\begin{eqnarray}
 T_{3,\ell}(n_\ell \minus 1,m_\ell|n_\ell,m_\ell) &=& \beta n_\ell, \nonumber\\
 T_{4,\ell}(n_\ell,m_\ell \minus 1|n_\ell,m_\ell) &=& \beta m_\ell. \label{eq:bio_transition2}
\end{eqnarray}
In addition to these reactions within a given cell, we allow for diffusion processes between neighboring sites. These are captured by the following reactions:
\begin{eqnarray}
 T_{5,\ell}(n_\ell \minus 1,n_{\ell'} \plus 1|n_\ell,n_{\ell'}) &=& D n_\ell \delta_{|\ell-\ell'|,1}, \nonumber\\
 T_{6,\ell}(m_\ell \minus 1,m_{\ell'} \plus 1|m_\ell,m_{\ell'}) &=& D m_\ell \delta_{|\ell-\ell'|,1}, \label{eq:bio_transition3}
\end{eqnarray}
where, again in the spirit of a minimalistic stylized model, we assume equal diffusion rates, $D$, for both substances.

\subsection{Linear-noise approximation and number of defects}
We proceed by carrying out the van Kampen analysis. On the deterministic level, one finds
\begin{subequations}
\begin{eqnarray}
 \dot{\psi}_\ell^\infty \!&=&\! 
  D\Delta\psi_\ell^\infty  +  \frac{1}{1 \plus \bigl[ g(t) \chi_\ell^\infty \bigr]^\gamma} - \beta \psi_\ell^\infty, \label{eq:bio_psi} \\
 \dot{\chi}_\ell^\infty \!&=&\! 
  D\Delta\chi_\ell^\infty  +  \frac{1}{1 \plus \bigl[ g(t) \psi_\ell^\infty \bigr]^\gamma} - \beta \chi_\ell^\infty. \label{eq:bio_chi}
\end{eqnarray}
\end{subequations}

If we again define the order parameter as ${\phi_\ell \equals (n_\ell \minus m_\ell)/V}$, and correspondingly ${\phi_\ell^\infty \equals \psi_\ell^\infty \minus \chi_\ell^\infty}$, then the central fixed point, ${\psi^*(g) \equals \chi^*(g)}$, corresponds to ${\phi_\ell^\infty \!\equiv\! \phi^* \equals 0}$, which is stable for ${g \lt g_b}$ and unstable for ${g \gt g_b}$.

As before, we make an adiabatic approximation and assume that, despite the fact that the control parameter $g$ is varied linearly in time, the dynamics operates near the symmetric deterministic fixed point $\psi^*(g)$ at all times, where ${g \equals g(t) \equals \mu t}$. These assumptions are valid up to the time, $t_c$, when the decision making occurs. As shown in Fig. \ref{fig:fig5}, the trajectories of the stochastic dynamics remain close to ${\psi^*(g) \equals \psi^*(\mu t)}$ up to that point. At next-to-leading order in the system-size expansion, we then find the following linear Langevin equation describing the fluctuations about the above deterministic dynamics:
\begin{equation}
 \dot{\lambda}_\ell = \left\{D\Delta - \beta\gamma\left[\beta\psi^*(\mu t)-1\right] - \beta\right\}\lambda_\ell + \eta_\ell(t).
\end{equation}
We have introduced $\lambda_\ell$ via the relation ${(n_\ell \minus m_\ell)/V \equals \phi_\ell^\infty+V^{-1/2}\lambda_\ell}$. The variables $\eta_\ell$ are Gaussian noise terms, and within the above adiabatic approximation, their variance and spatial correlations are dependent on the value of the central fixed point,
\begin{eqnarray}
 \avg{\eta_\ell(t)\eta_{\ell'}(t')} \!&=&\! \delta(t-t')\bigl[(4\beta+8D)\psi^*(\mu t)\delta_{\ell,\ell'}\nonumber\\
  && \qquad\quad\;\; -4D\psi^*(\mu t)\delta_{|\ell-\ell'|,1}\bigr]. \label{eq:bio_noise}
\end{eqnarray}
At the symmetric deterministic fixed point, we have ${\phi^* \equals 0}$, and linearization about this value gives ${\phi_\ell \equals V^{-1/2}\lambda_\ell}$. Within the linear approximation, we can therefore write
\begin{equation}
 \dot{\phi}_\ell = \left\{D\Delta - \beta \gamma\left[{\beta} \psi^*(\mu t)\minus1\right] - \beta\right\}\phi_\ell + V^{-1/2}\eta_\ell(t). \label{eq:bio_phi}
\end{equation}

Following the previous section, we can obtain a closed-form solution for $\avg{\phi^2(t)}$, reported in more detail in Appendix \ref{app:bio}. This expression can be integrated numerically, and used to find ${\hat{g} \equals g_c \minus g_b}$, where ${g_c \equals g(t_c)}$, and where the time of the freeze-out, $t_c$, is obtained from 
\begin{equation}
 \avg{\phi^2(t_c)} = \delta \bigl[\phi^*(t_c)\bigr]^2. \label{eq:bio_ghat}
\end{equation}
The quantity $\phi^*(t)$ is the stable non-zero fixed point for ${g(t) \gt g_b}$. To calculate the expected density of zeros, we use Eq. (\ref{eq:numzeros_gamma}), where $\Gamma(t_c)$ is the width of the structure factor evaluated at the freeze-out time defined by Eq. (\ref{eq:bio_ghat}).

\subsection{Test against simulations}
We test these analytical predictions against simulations of the process defined by Eqs.  (\ref{eq:bio_transition1}), (\ref{eq:bio_transition2}), and (\ref{eq:bio_transition3}) in Fig. \ref{fig:fig6}. Again, as seen in Secs. \ref{sec:opinion} and \ref{sec:evol}, we find good agreement between simulations and theory for the quantity $\hat{g}$; see the inset of the figure. The zero crossings of the order-parameter field are counted subject to a minimum size threshold defined by one-tenth of the rms field amplitude, as discussed in Sec. \ref{sec:opinion}. 
In the main panel of the figure, it is seen that for fast quenches these values agree with the theoretical prediction, however for slow quenches the density of defects deviates from the prediction. We attribute this to difficulties in counting zeros when ${g_c \!\sim\! g_b}$. The measured width of the structure factor shows good agreement with the theory for all quench rate values, even when the density of defects shows deviations. As explained above, we consider this width the more fundamental quantity throughout this work. We note, however, that deviations between theory and simulations are stronger for this model than for those of the previous sections. This is presumably due to stronger nonlinearities in the cell decision model. Nevertheless, our results show that the theory based on the linear-noise approximation can successfully predict the delay of the bifurcation and that it captures the characteristic length scale of the resulting patterns to a good degree.

\begin{figure}[t]\includegraphics[width=\columnwidth]{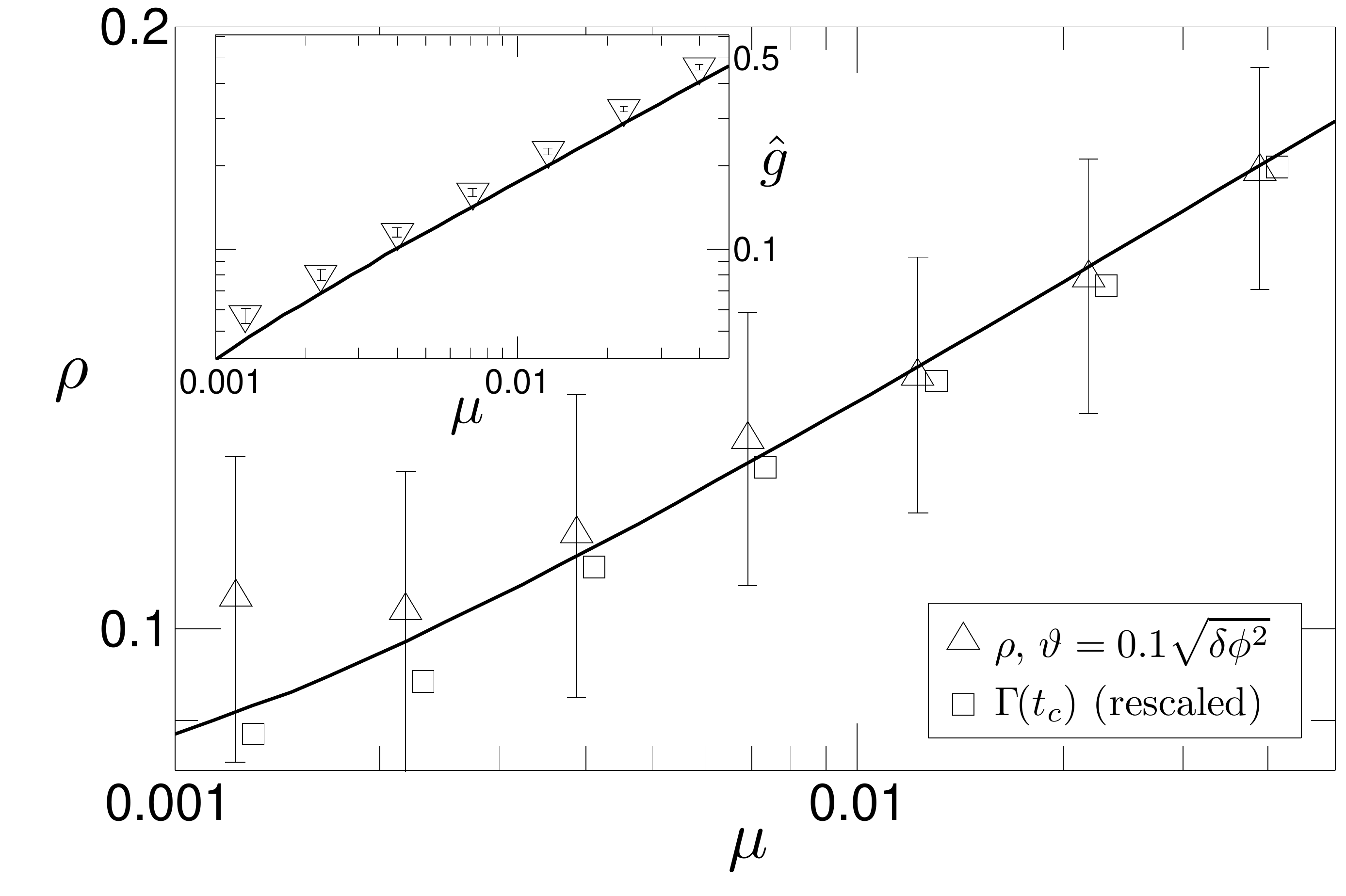}
\caption{Density of defects in the model of cellular decision making at ${g \equals g_c}$. Triangles show results from numerical simulations of the individual-based model, Eqs. (\ref{eq:bio_transition1}), (\ref{eq:bio_transition2}), and (\ref{eq:bio_transition3}). As in the previous models, a threshold is applied when counting zero crossings. Squares show the width of the structure factor, $\Gamma(t_c)$, measured in simulations, rescaled to collapse with $\avg{\rho}$. The solid line shows the theoretical predictions, obtained using Eq. (\ref{eq:numzeros_gamma}), and the expression for the structure factor given in Appendix \ref{app:bio}. Inset: Values of $\hat{g}$ from stochastic simulations. The solid line corresponds to the solution of Eq. (\ref{eq:bio_ghat}). Error bars represent standard deviations over $100$ realizations. Model parameters are as in Fig. \ref{fig:fig5}.}
\label{fig:fig6}
\end{figure}

\section{Growing populations}\label{sec:growing}
\subsection{Background and model definition}
The pattern-forming processes in the models investigated in the previous sections are due to a gradual change in the underlying potential of the deterministic limiting dynamics. While these potentials all have a single minimum at the beginning of the sweep, a double-well regime is entered, subsequently triggering the symmetry-breaking. It is this change of potential that brings about the defect formation. In this section, we will consider a different pattern-forming mechanism, and we focus on models in which the external parameters remain fixed in the symmetry-broken phase, ${g \gt g_b}$, i.e., when the limiting deterministic dynamics has multiple stable fixed points. In the model discussed in this section, the symmetry-breaking is instead triggered by a gradually decreasing noise amplitude, originating from persistent growth of the population. Specifically, we will consider an exponential growth process of the overall population size, $N(t)$, leading to a decreasing amplitude of the resulting demographic fluctuations, which scale as $N^{-1/2}$. At the beginning of the dynamics, when the population is small, fluctuations will be large, hence masking the double-well structure of the deterministic dynamics. The system remains in a disordered state. As the population grows and fluctuations become smaller, the deterministic drift will become increasingly relevant. When the noise amplitude is of the same order as the separation of the deterministic fixed points finally, the dynamics locally (i.e. in each lattice site) choose one of the two equilibrium fixed points, and local population numbers fluctuate about these symmetry-broken equilibria. 

Specifically, we will consider the model of opinion dynamics, defined in Eq. (\ref{eq:op_transition}) in Sec. \ref{sec:opinion}, where $g$ is now constant in time. The growth dynamics is introduced by two additional reactions,

\begin{eqnarray}
 T_{5,\ell}(n_\ell \plus 1,m_\ell|n_\ell,n_\ell) \!&=&\! \frac{\mu}{2}N_\ell, \nonumber\\
 T_{6,\ell}(n_\ell, m_\ell \plus 1|n_\ell,m_\ell) \!&=&\! \frac{\mu}{2}N_\ell, \label{eq:grow_transition}
\end{eqnarray}
where ${N_\ell \equals n_\ell \plus m_\ell}$ is the total population in lattice site $\ell$, and where $\mu$ is the growth rate. We have written $n_\ell$ for the number of up-spins in lattice site $\ell$, and $m_\ell$ for the number of down-spins. At any one time, one has ${N_\ell \equals n_\ell \plus m_\ell}$. We note that the offspring created are randomly assigned to either type of individual (up- and down-spins). Making a deterministic approximation, we find
\begin{eqnarray}
 \frac{d}{dt} \avg{n_\ell} &=& 
  \frac{\avg{N_\ell}}{2}\left\{D\Delta \phi_\ell^\infty 
  +a\phi_\ell^\infty \bigl[g-(\phi_\ell^\infty)^2\bigr] + \mu \right\}, \nonumber \\
 \frac{d}{dt}\avg{m_\ell} &=& 
  -\frac{\avg{N_\ell}}{2}\left\{D\Delta \phi_\ell^\infty
  +a\phi_\ell^\infty \bigl[g-(\phi_\ell^\infty)^2\bigr] - \mu\right\}, \nonumber \\
\end{eqnarray}
where
\begin{equation}
 \avg{N_\ell(t)} = N_0 e^{\mu t},
\end{equation}
and where we have written ${\phi_\ell^\infty(t) \equals \avg{n_\ell(t) \minus m_\ell(t)}/\avg{N_\ell(t)}}$. This order-parameter field in turn follows the deterministic dynamics,
\begin{equation}
 \dot{\phi}_\ell^\infty = D\Delta \phi_\ell^\infty + a\phi_\ell^\infty \bigl[g-(\phi_\ell^\infty)^2\bigr] - \mu\phi_\ell^\infty.
\end{equation}
The spatially homogeneous fixed points of this dynamics are given by ${\phi^* \equals \pm \sqrt{g \minus \mu/a}}$. In our simulations, we study the stochastic dynamics for different values of the growth rate, $\mu$. To keep the deterministic fixed points at a fixed location, we adjust the value of $g$ such that ${g \minus \mu/a \equals 0.5}$ in all simulation runs.

\begin{figure}[t]\includegraphics[width=1.0\columnwidth]{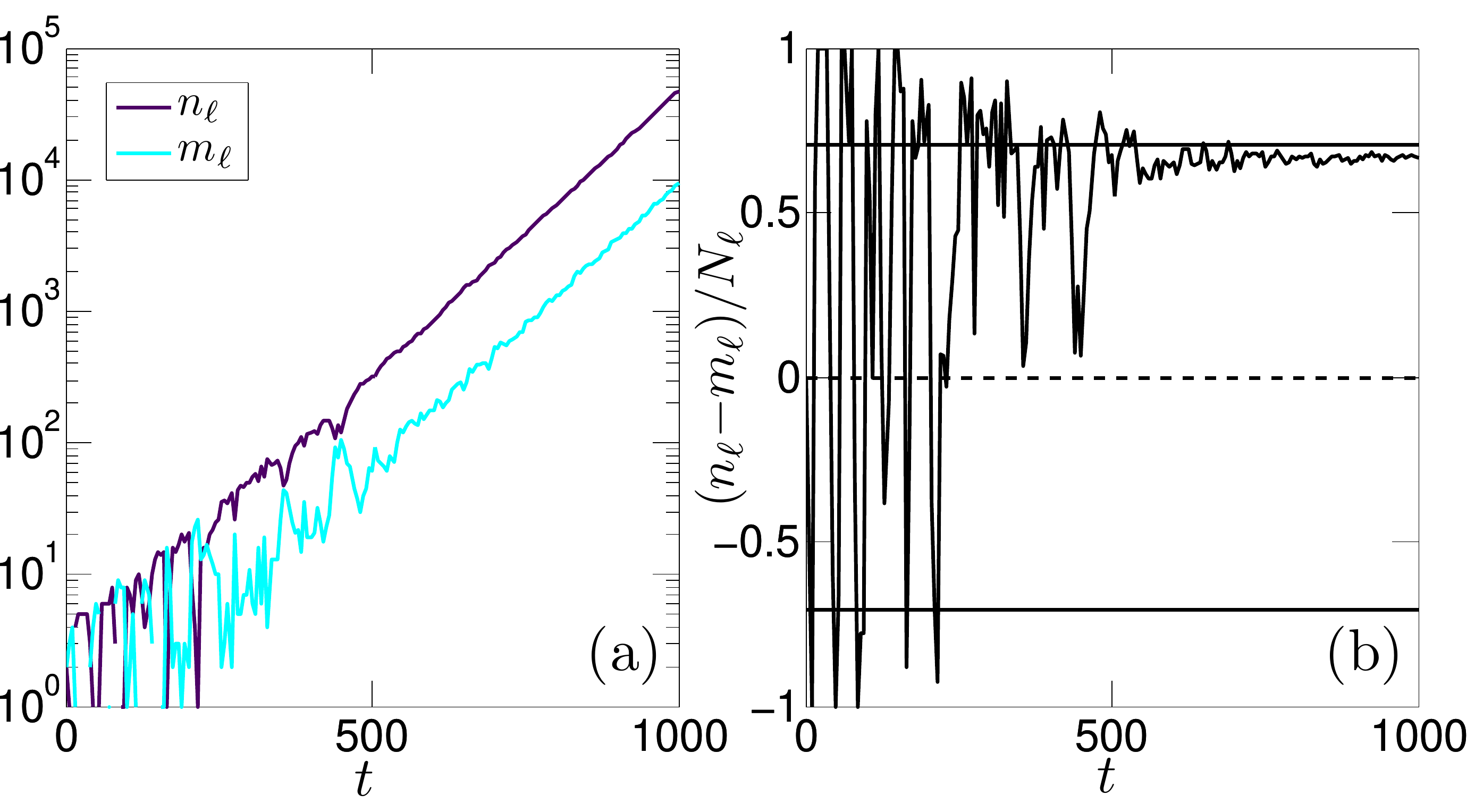}
\caption{(Colour on-line) (a): Number of up-spins and down-spins in a growing population at a fixed lattice site in the model of opinion dynamics. (b): Order parameter, $\phi_\ell(t)$. Data are from a single run with model parameters ${\mu \equals 0.01}$, ${a \equals 0.25}$, ${g \minus \mu/a \equals 0.5}$, ${D \equals 0.05}$, ${N_0 \equals 4}$, and ${L \equals 200}$.}
\label{fig:fig7}
\end{figure}

\subsection{Simulation results}

The dynamics of the stochastic model are illustrated in Fig. \ref{fig:fig7}. Figure \ref{fig:fig7}(a) shows the particle numbers, $n_\ell$ and $m_\ell$, at a single lattice site. As seen in the figure, they grow exponentially. At the beginning of the dynamics, the population is small, and so fluctuations are large. The noise settles down during the later parts, and in this particular run the population of up-spins outgrows the population of down-spins. This is seen in Fig. \ref{fig:fig7}(b). The order parameter ${\phi_\ell \equals (n_\ell \minus m_\ell)/N_\ell}$ is subject to large fluctuations at the beginning. The amplitude of these fluctuations is sufficiently large to make the attractors of the deterministic dynamics largely irrelevant initially. In the later stages the noise is smaller, and the run of the stochastic dynamics shown in the figure approaches values near the positive fixed point of the deterministic dynamics.

Examples of the resulting patterns in the spatial system are shown in Fig. \ref{fig:fig8}. As in the previous examples, the typical size of the resulting domains depends on the quench rate. For fast quenches, fine structures with a large number of defects are obtained, while larger domains emerge in slow quenches. We stress again that these patterns were generated at a constant control parameter The role of the quench rate here is taken by the growth rate, $\mu$, i.e., the rate with which the noise amplitude is reduced over time. The number of zero crossings of the order-parameter field is counted at the end of the simulation (${t \equals 10/\mu}$). No threshold is applied when counting zeros of the field as the defects are well formed at the end of the simulation; as seen in Fig. \ref{fig:fig8}. Fitting the number of zeros as a function of the growth rate to a power law gives an exponent of approximately $0.23$; see Fig. \ref{fig:fig9}. While this is close to the KZ prediction for systems in the class of the Ginzburg-Landau equation, we have no physical justification for why the KZ theory should apply here. It is clear, however, that the size of the domains that are formed depends on the growth rate, $\mu$. We speculate that this may have implications for pattern-forming processes, for example in growing embryos, or indeed in other systems in evolutionary dynamics with exponentially growing populations \cite{MCF10,CMF12}.  

\begin{figure}[t]\includegraphics[width=\columnwidth]{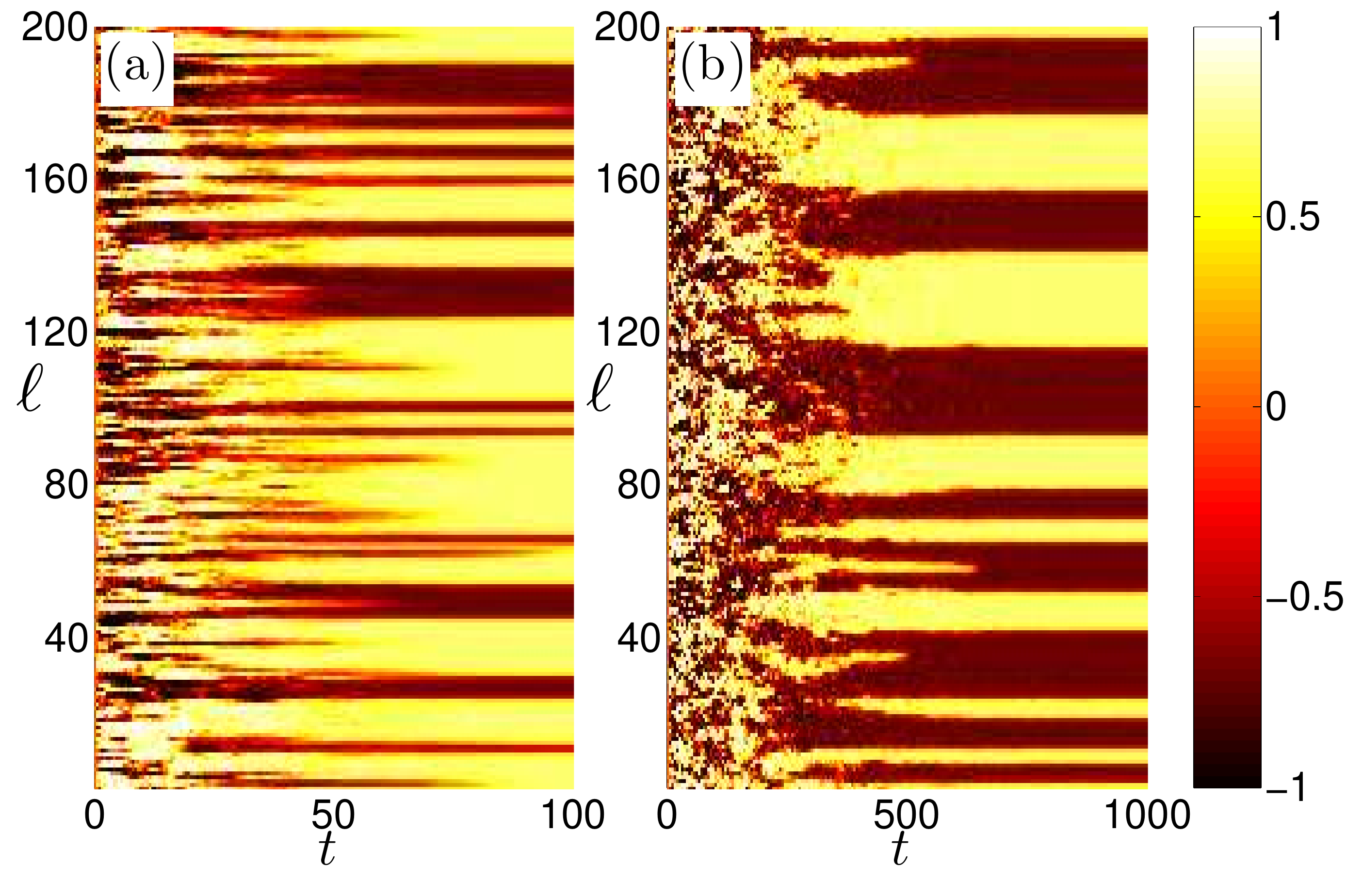}
\caption{(Color on-line). Spatio-temporal dynamics of the order parameter, $\phi_\ell(t)$, for the model of opinion dynamics with growing populations in each lattice site. Light shading indicates high values of $\phi_\ell(t)$, dark shading indicates low values. Panel (a) shows data from a single simulation run at ${\mu \equals 0.10}$, panel (b) is for ${\mu \equals 0.01}$. The remaining parameters are as in Fig. \ref{fig:fig7}.}
\label{fig:fig8}
\end{figure}

\begin{figure}[t]\includegraphics[width=\columnwidth]{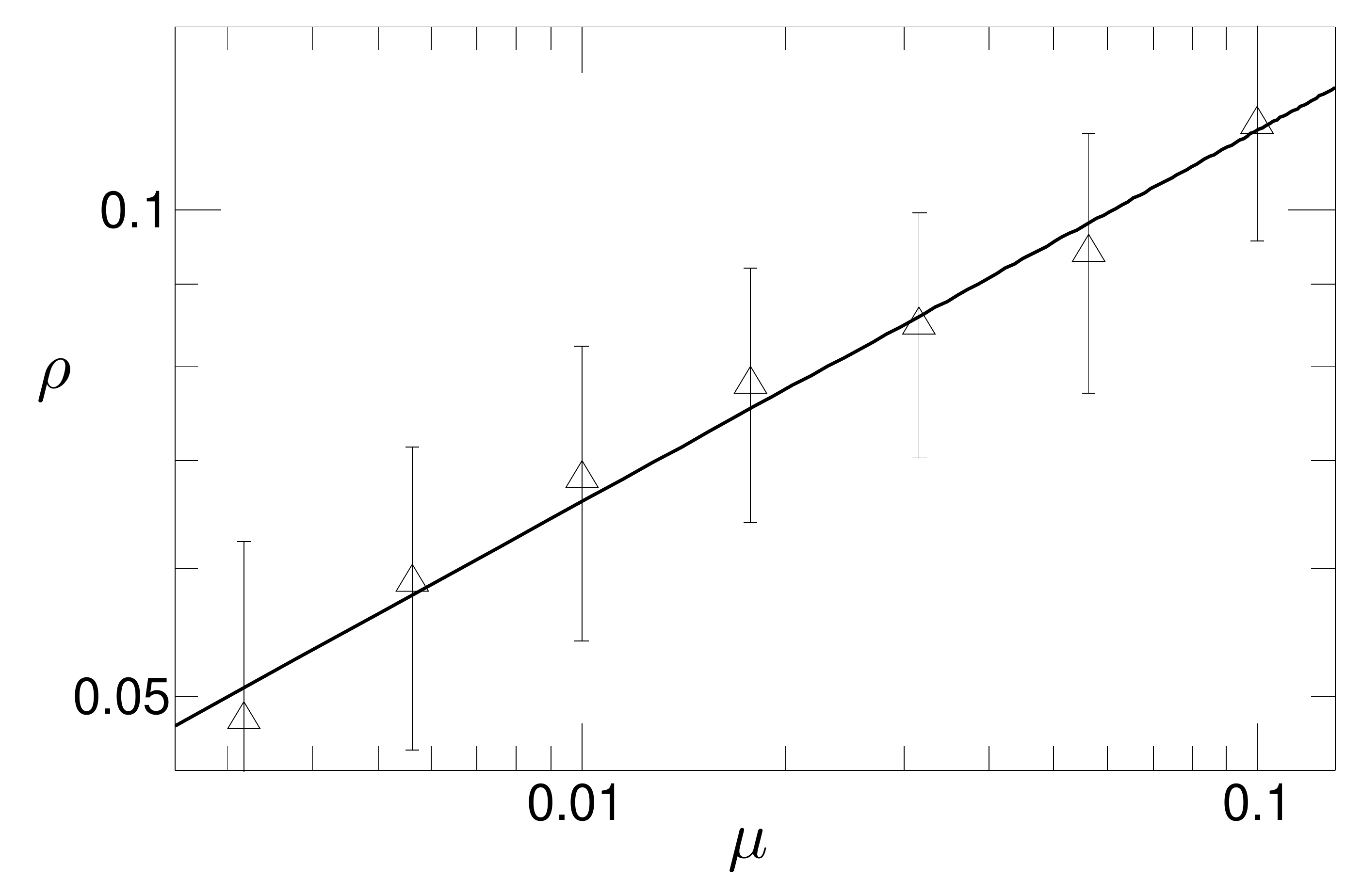}
\caption{Number of zeros of the order-parameter field at ${t \equals 10/\mu}$ in the model of opinion dynamics with growing populations. Symbols show data from numerical simulations, averaged over $100$ samples. Error bars indicate the resulting standard deviation. Model parameters are as in Fig. \ref{fig:fig7}. The solid line is a least-squares fit to a power law, resulting in an exponent of $0.23$.}
\label{fig:fig9}
\end{figure}

\section{Conclusions}\label{sec:concl}
In this paper, we have extended the picture of pattern formation in individual-based models to include systems with time-dependent parameters. Similar to what is known as the Kibble-Zurek mechanism, the length scale of the resulting patterns depends on the rate with which the system is swept across its symmetry-breaking transition. While most existing studies of such phenomena are based on systems to which external noise is added to partial differential equations with time-dependent parameters, we focus here on intrinsic noise, originating from the discrete dynamics at the microscopic level. Thus, while the resulting phenomenology is similar to what is known in condensed matter systems, the source of the noise is different from that in existing models. Where possible, we make use of linear-noise approximations to derive analytical approximations for the characteristic length scale of spatial structures and for the density of defects resulting from the finite-time quenches.  

Our analysis demonstrates that the picture of defect formation in systems with time-varying parameters is applicable in a number of different model systems. In particular, we have looked at a simple model of opinion dynamics, designed to reduce to the well-known Ginzburg-Landau equation in the linear-noise limit. Our second exemplar is a model of selection-mutation dynamics in the context of evolutionary game theory. The time-varying element here is the payoff structure of the underpinning game, which gradually evolves from a co-existence game to a co-ordination game. Our third example is a model of decision making in biology, describing two fate-determining chemical substances with mutual inhibition, and a time-dependent interaction coefficient, driving the system from an undifferentiated to a differentiated state. In all of these models, fast quenches lead to small-scale patterns with a large number of defects separating domains. Slower quenches, on the other hand, generate large domains, with relatively few defects. The characteristic length scale of these patterns can be approximated successfully in all cases. 

In our final example, we have considered a separate, noise-driven mechanism of pattern formation. In this model we consider a growing population, so that the magnitude of demographic fluctuations decreases with time. While the large amplitude of the noise masks the underlying deterministic drift in the initial phases of the dynamics, the system is driven towards the deterministic attractors as the noise is reduced. Our simulations demonstrate that the size of domains in which the same attractor is chosen scales with the growth rate of the population. For fast growth we find small domain sizes; for slow growth there is sufficient time for information to travel through the system and for different spatial locations to coordinate on the same attractor. As a result, only relatively few defects emerge. 

In summary, our analysis shows that the time scales on which model parameters such as reaction rates or population sizes change in individual-based systems may crucially affect the spatial structures that these systems generate. This is a combined effect of an underlying symmetry-breaking, delayed bifurcation dynamics induced by time-varying model parameters, and intrinsic noise triggering the symmetry breaking. While we have focused here on a set of relatively stylized models, we expect the basic phenomenology to be relevant in a variety of biological systems subject to external time-dependent signals and to internal fluctuations. Our work makes a connection with what has been known about condensed matter systems swept across symmetry-breaking transitions. We show how existing tools, combined with a linear-noise approximation, can be used to predict the properties of patterns generated by individual-based systems with time-varying parameters.

\medskip

{\em Acknowledgements.} P.A. acknowledges support from the EPSRC. T.G. would like to thank Esteban Moro for an earlier collaboration on the dynamics of defect formation.

\appendix
\section{Further details of the linear-noise approximation for the model of evolutionary dynamics}\label{app:noise}
In this appendix, we present further details of the system-size expansion and the linear-noise approximation for the model discussed in Sec. \ref{sec:evol}. In order to study the effects of fluctuations we follow the same process as in Sec. \ref{sec:opinion} and derive a Fokker-Planck equation for the probability distribution of the variables ${(\vec{\xi},\vec{\zeta})}$, and from this an equivalent set of Langevin equations. We find
\small
\begin{equation}
 \begin{pmatrix}
  \dot{\xi}_\ell \\ \dot{\zeta}_\ell
 \end{pmatrix}
 =
 \mathcal{J}(t)
 \begin{pmatrix}
  \xi_\ell \\ \zeta_\ell
 \end{pmatrix}
 +
 \begin{pmatrix}
  \eta_{A,\ell}(t) \\ \eta_{B,\ell}(t)
 \end{pmatrix},\label{eq:evo_fluc}
\end{equation}
\normalsize
where $\mathcal{J}(t)$ is the Jacobian of Eqs. (\ref{eq:evo_psi}) and (\ref{eq:evo_chi}) to be evaluated on the deterministic trajectory. Our approach assumes that the dynamics operates near the symmetric deterministic fixed point, and so we use ${\psi_\ell^\infty \equals \chi_\ell^\infty \equals 1/2}$. The quantities $\eta_{A,\ell}$ and $\eta_{B,\ell}$ are Gaussian noise variables, with correlations across components of the form  
\begin{equation}
 \avg{\eta_{i,\ell}(t)\eta_{j,\ell'}(t')} = \mathcal{B}_{ij}^{(\ell,\ell')} \delta(t-t')
\end{equation}
where ${i,j \!\in\!\{A,B\}}$. As we only consider nearest-neighbor interactions, non-zero entries of $\mathcal{B}^{(\ell,\ell')}$ only occur when ${|\ell-\ell'| \les 1}$. Using this, we can write 
\begin{equation}
 \mathcal{B}^{(\ell,\ell')} = b^{(0)}\delta_{|\ell-\ell'|,0} + b^{(1)}\delta_{|\ell-\ell'|,1}
\end{equation}
where the matrices $b^{(0)}$ and $b^{(1)}$ are given by
\small
\begin{equation}
 b^{(0)} = \begin{pmatrix}
            \frac{1}{4} \plus 2D & -\frac{1}{4} \\
	    -\frac{1}{4} & \frac{1}{4} \plus 2D
           \end{pmatrix},
 \quad
 b^{(1)} = \begin{pmatrix}
            -D & 0 \\
	    0 & -D
           \end{pmatrix}.
\end{equation}
\normalsize

For the Langevin equation (\ref{eq:evo_lambda}) for the variable ${\lambda_\ell \equals \xi_\ell \minus \zeta_\ell}$, the noise term satisfies ${\eta_\ell \equals \eta_{A,\ell} \minus \eta_{B,\ell}}$, and thus the correlator is found to be
\small
\begin{eqnarray}
 \avg{\eta_\ell(t)\eta_{\ell'}(t')} \!&=&\! 
  \avg{\eta_{A,\ell}(t)\eta_{A,\ell'}(t')} - \avg{\eta_{A,\ell}(t)\eta_{B,\ell'}(t')} \nonumber\\
 &&\! -\avg{\eta_{B,\ell}(t)\eta_{A,\ell'}(t')} + \avg{\eta_{B,\ell}(t)\eta_{B,\ell'}(t')},\nonumber\\
\end{eqnarray}
\normalsize
which gives Eq. (\ref{eq:evo_noise}). One can now take the spatial Fourier transform  with respect to the variable ${\ell \minus \ell'}$, and find
\begin{equation}
 \avg{\tilde{\eta}_q(t)\tilde{\eta}_q(t')} = \frac{1}{2\pi}\delta(t-t')\left[1+4D(1-\cos q)\right]. 
\end{equation}

\section{Further details of the calculation of $\avg{\phi^2(t)}$ for the model of cell decision making}\label{app:bio}
In Sec. \ref{sec:cells} we arrive at the Langevin equation (\ref{eq:bio_phi}), which describes the evolution of the order parameter when linearized about the ${\phi^* \equals 0}$ fixed point, which represents the equal-concentration fixed point ${\psi_\ell^\infty \equals \chi_\ell^\infty \equals \psi^*(g)}$. The correlator of the noise in the Langevin equation is given by Eq. (\ref{eq:bio_noise}). Carrying out a Fourier transform of the Langevin equation (with respect to the spatial variable, $\ell$) gives
\begin{equation}
\dot{\tilde{\phi}}_q = -\left\{Dq^2 + \beta \gamma\left[{\beta} \psi^*(t)\minus1\right] + \beta\right\}\tilde{\phi}_q + V^{-1/2}\tilde{\eta}_q(t),
\end{equation}
where the correlations of the Fourier components of the noise are given by
\begin{eqnarray}
 \avg{\tilde{\eta}_q(t)\tilde{\eta}_{q'}(t')} 
  &=& \frac{1}{2\pi}\delta(t-t')\delta(q+q')\nonumber \\
  && \hspace{-5em}\times \left[4\beta+8D(1-\cos q)\right]\psi^*(\mu t). 
\end{eqnarray}
From this, the structure factor is calculated as
\begin{eqnarray}
 S(q,t) &=& \frac{4\beta+8D(1-\cos q)}{2\pi V} \nonumber \\
 && \hspace{-5em}\times e^{-2[Dq^2 + \beta(1-\gamma)]t -2\beta^2\gamma\int^t \!ds\,\psi^*(\mu s)} \nonumber\\
&& \hspace{-5em} \times \int_{t_0}^t dt'\;\psi^*(\mu t')e^{2[Dq^2 + \beta(1-\gamma)]t' +2\beta^2\gamma\int^{t'} \!\!ds\,\psi^*(\mu s)}, \nonumber\\ \label{eq:bio_sf}
\end{eqnarray}
and $\avg{\phi^2(t)}$ is defined by the integral of $S(q,t)$ over the Fourier variable $q$.


\begin{thebibliography}{99}
 \bibitem{AT52} A. M. Turing, Philos. Trans. R. Soc. London, Ser. B {\bf 237}, 37 (1952).

 \bibitem{CG09} M. C. Cross and H. S. Greenside, {\em Pattern Formation and Dynamics in Non-Equilibrium Systems} (Cambridge University Press, Cambridge, 2009).
    

 \bibitem{Kibble76} T. W. B. Kibble, J. Phys. A {\bf 9}, 1387 (1976).

 \bibitem{Zurek85} W. H. Zurek, Nature (London) {\bf 317}, 505 (1985).

 \bibitem{LZ97} P. Laguna and W. H. Zurek, Phys. Rev. Lett. {\bf 78}, 2519 (1997).

 \bibitem{YZ98} A. Yates and W. H. Zurek, Phys. Rev. Lett. {\bf 80}, 5477 (1998).

 \bibitem{lythe} G. D. Lythe, Phys. Rev. E {\bf 53}, R4271 (1996).

 \bibitem{ML99} E. Moro and G. D. Lythe, Phys. Rev. E {\bf 59}, R1303 (1999).

 \bibitem{GM03} T. Galla and E. Moro, Phys. Rev. E {\bf 67}, 035101 (2003).


 \bibitem{LCD} I. Chuang, R. D\"urrer et. al., Science {\bf 251}, 1336 (1991).
 
 \bibitem{LCD2} M. J. Bowick, L. Chandar et. al., Science {\bf 263}, 943 (1994).

 \bibitem{Neutron} C. B\"auerle, Y. M. Bunkov et. al., Nature (London) {\bf 382}, 332 (1996).
 
 \bibitem{Neutron2} V. H. M. Ruutu, V. B. Elstov et. al., Nature (London) {\bf 382}, 334 (1996).

 \bibitem{DR99} S. Ducci, P.L. Ramazza, W. Gonzalez-Vinas, and F. T. Arecchi, Phys. Rev. Lett. {\bf 83}, 5210 (1999).

 \bibitem{CL12} S. C. Chae, N. Lee et. al., Phys. Rev. Lett. {\bf 108}, 167603 (2012).

 \bibitem{hendry} P. C. Hendry, N. S Lawson et. al., Nature (London) {\bf 368}, 315 (1994).

 \bibitem{ulm} S. Ulm, J. Ro\ss nagel et. al., Nat. Commun. {\bf 4}, 2290 (2013).

 \bibitem{casado} S. Casado, W. Gonzalez-Vinas et al., Eur. Phys. J. Spec. Top. {\bf 146}, 87 (2007).

 
 \bibitem{MN05} A. J. McKane and T. J. Newman, Phys. Rev. Lett. {\bf 94}, 218102 (2005).
 
 \bibitem{RMF06} T. Reichenbach, M. Mobilia, and E. Frey, Phys. Rev. E {\bf 74}, 051907 (2006).

 \bibitem{KGG07} R. Kuske, L.~F. Gordillo, and P. Greenwood, J. Theor. Biol. {\bf 245}, 459 (2007).
 
 \bibitem{PBD07} M. Pineda-Krch, H. J. Blok, U. Dieckmann et. al., OIKOS {\bf 116}, 53 (2007).
 
 \bibitem{AMP07} D. Alonso, A.~J. McKane, and M. Pascual, J. Roy. Soc. Interface {\bf 4}, 575 (2007).

 \bibitem{STN08} M. Simoes, M.M. Telo~da~Gama, and A. Nunes, J. Roy. Soc. Interface {\bf 5}, 555 (2008).


 \bibitem{BG11} T. Butler and N. Goldenfeld, Phys. Rev. E {\bf 84}, 011112 (2011).

 \bibitem{BF10} T. Biancalani, D. Fanelli, and F. Di Patti, Phys. Rev. E {\bf 81}, 046215 (2010).


 \bibitem{LM08} C. A. Lugo and A. J. McKane, Phys. Rev. E {\bf 78}, 051911 (2008).

 \bibitem{SPT11} M. Scott, F. J. Poulin, and H. Tang, Proc. R. Soc. London Ser. A {\bf 467}, 718 (2011).
 
 \bibitem{BGM11} T. Biancalani, T. Galla, and A. J. McKane, Phys. Rev. E {\bf 84}, 026201 (2011).


 \bibitem{leibler1} E. Kussell and S. Leibler, Science {\bf 309}, 2075 (2005).
 
 \bibitem{leibler2} S. Leibler and E. Kussell, Proc. Natl. Acad. Sci. (U.S.A.) {\bf 107}, 13183 (2010).

 \bibitem{alexei1} N. R. Nen\'e, J. Garca-Ojalvo, and A. Zaikin, PLoS ONE {\bf 7}, e32779 (2012).

 \bibitem{alexei2} N. R. Nen\'e and A. Zaikin, PLoS ONE {\bf 7}, e40085 (2012).

 \bibitem{gerland} U. Gerland, T. Hwa, Proc. Natl. Acad. Sci. (U.S.A.) {\bf 106}, 8841 (2009).

 \bibitem{waddington} C. H. Waddington, {\em The Strategy of the Genes}, (Allen \& Unwin, London, 1957).
 
 
 \bibitem{RB11} D. I. Russell and R. A. Blythe, Phys. Rev. Lett. {\bf 106}, 165702 (2011).
 
 \bibitem{antunes06} N. D. Antunes, P. Gandra, and R. J. Rivers, Phys. Rev. D {\bf 73}, 125003 (2006).

 \bibitem{biroli10} G. Biroli, L. F. Cugliandolo, and A. Sicilia, Phys. Rev. E {\bf 81}, 050101 (2010).

 \bibitem{chandran12} A. Chandran, A. Erez, S. S. Gubser, and S. L. Sondhi, Phys. Rev. B {\bf 86}, 064304 (2012).

 \bibitem{remark:defects} We note that there has been some discussion as to whether the density of defects is set before or after $g_b$ is crossed during the quench; see \cite{antunes06,biroli10} for recent discussions. The phenomenological details of Zurek's picture regarding this point are not immediately relevant for our work, as we are able to derive analytical approximations for the density of defects from a linear-noise approximation, as discussed below.

 \bibitem{DS01} J. Dziarmaga and M. Sadzikowski, Phys. Rev. E {\bf 63}, 036112 (2001).

 \bibitem{D05} J. Dziarmaga, Phys. Rev. Lett. {\bf 95}, 245701 (2005).


 \bibitem{castellano} C. Castellano, S. Fortunato, and V. Loreto, Rev. Mod. Phys. {\bf 81}, 591 (2009).


 \bibitem{vK} N.G. van Kampen, {\em Stochastic Processes in Physics and Chemistry}, 3rd ed. (Elsevier, Amsterdam, 2007).


 \bibitem{parseval} M. A. Parseval de Ch\^enes, Sci. Math. Phys.(Savans \'Etrangers) {\bf 1}, 638 (1806).


 \bibitem{LHM} B. I. Halperin, in {\em Physics of Defects}, Proceedings of Les Houches Summer School, Session XXXV, edited by R. Balian, M. K. Kl\'eman, and J. P. Poirier (North-Holland Press, Amsterdam, 1981), p. 816.
 
 \bibitem{LHM2}  F. Liu and G. F. Mazenko, Phys. Rev. B {\bf 46}, 5963 (1992).

 \bibitem{gillespie} D. T. Gillespie, J. Phys. Chem. {\bf 81}, 2340 (1977).
 
 \bibitem{remark:gillespie} As the reaction rates in this model are explicitly time-dependent, the stochastic simulation algorithm does not simulate the master equation exactly. Instead, at each simulation step, the rates are fixed using the values at the beginning of the step. The step size is of order $(NL)^{-1}$, so we expect any resulting errors to be irrelevant to the order of the system-size expansion at which we are working.

 \bibitem{LytheThesis} G. D. Lythe, Ph.D. thesis, University of Cambridge, (1994).


 \bibitem{hauert} A. Traulsen and C. Hauert, {\em Stochastic Evolutionary Game Dynamics}, in Vol. 2 of Reviews of Nonlinear Dynamics and Complexity, edited by H. G. Schuster, (Wiley-VCH, Weinheim, Germany, 2010).

 \bibitem{TCH05} A. Traulsen, J. C. Claussen, and C. Hauert, Phys. Rev. Lett. {\bf 95}, 238701 (2005).

 \bibitem{remark:beta} Model parameters must be chosen such that ${1\plus \beta(\Pi_A\minus\Pi_B) \equals 1 \minus \beta g(t)\phi_\ell > 0}$ for all $g(t)$. In principle, $\phi_\ell$ can fluctuate by arbitrary amounts. In our simulations parameters, are chosen such that the above condition is fulfilled for all ${|\phi| \lt 1+4\Omega^{-1/2}}$. Fluctuations of $\phi_\ell$ of more than $4\Omega^{-1/2}$ are sufficiently rare not to occur in practice.

 \bibitem{JIM12} J. Jaeger, D. Irons, and N. Monk, J. Exp. Zool. (Mol. Dev. Evol.) {\bf 318B}, 591 (2012).

 \bibitem{GCC00} T. S. Gardner, C. R. Cantor, and J. J. Collins, Nature (London) {\bf 403}, 339 (2000).

 \bibitem{Ferrell12} J. E. Ferrell, Curr. Bio. {\bf 22}, 458 (2012).

 \bibitem{hill} A. V. Hill, J. Physiol. {\bf 40}, iv (1910).
  

 \bibitem{MCF10} A. Melbinger, J. Cremer, and E. Frey, Phys. Rev. Lett. {\bf 105}, 178101 (2010).

 \bibitem{CMF12} J. Cremer, A. Melbinger, and E. Frey, Sci. Rep. {\bf 2}, 281 (2012).
\end{thebibliography}
\end{document}